\documentclass[%
 aip,
 amsmath,amssymb,
 reprint,%
nofootinbib]{revtex4-1}

\usepackage{graphicx}
\usepackage{dcolumn}
\usepackage{bm}
\usepackage[usenames]{color}
\usepackage[dvipsnames]{xcolor}
\usepackage[utf8]{inputenc}
\usepackage[T1]{fontenc}
\usepackage{mathptmx}
\usepackage{subcaption}
\usepackage{soul}
\begin{document}

\preprint{AIP/123-QED}

\title{Dynamics of multilayer networks with amplification}

\author{Thierry Njougouo}
 \affiliation{Research Unit Condensed Matter, Electronics and Signal Processing,
University of Dschang, P.O. Box 67 Dschang, Cameroon.}

\author{Victor Camargo}%
\affiliation{Center for Interdisciplinary Research on Complex Systems,University
of Sao Paulo, Av. Arlindo Bettio 1000, 03828-000 S\~ao Paulo, Brazil.}%

\affiliation{Department of Physics-FFCLRP, University of S\~ao Paulo, Ribeirao
Preto-SP, 14040-901, Brasil.}

\author{Patrick Louodop}
\affiliation{Research Unit Condensed Matter, Electronics and Signal Processing,
University of Dschang, P.O. Box 67 Dschang, Cameroon.}%

\author{Fernando Fagundes Ferreira}%
\affiliation{Center for Interdisciplinary Research on Complex Systems,University
of Sao Paulo, Av. Arlindo Bettio 1000, 03828-000 S\~ao Paulo, Brazil.}%

\affiliation{Department of Physics-FFCLRP, University of S\~ao Paulo, Ribeirao
Preto-SP, 14040-901, Brasil.}

\author{ Pierre K. Talla}%
\affiliation{L2MSP, University of Dschang, P.O. Box 67 Dschang, Cameroon.}%

\author{ Hilda A. Cerdeira}%
\affiliation{S\~ao Paulo State University (UNESP), Instituto de F\'{i}sica
Te\'{o}rica, Rua Dr. Bento Teobaldo Ferraz 271, Bloco II, Barra Funda, 01140-070
S\~ao Paulo, Brazil.}%

\date{\today}

\begin{abstract}
	We study the dynamics of a multilayer network of chaotic oscillators
	subject to an amplification. Previous studies have proven that  multilayer
	networks present phenomena such as synchronization, cluster and chimera states.
	Here we consider a network with two layers of
	R\"ossler chaotic oscillators as well as
	applications to multilayer networks of chaotic jerk and
	Li\'enard oscillators. Intra-layer coupling is considered to be all to all
	in the case of R\"ossler oscillators, a ring for jerk oscillators and
	global mean field coupling in the case of Li\'enard, the inter-layer coupling is
	unidirectional in all these three cases. The second layer has
	an amplification coefficient. An in-depth study  on the case of a
	network of R\"ossler oscillators using master stability
	function and order parameter leads to several phenomena such as complete
	synchronization, generalized, cluster
	and phase synchronization with amplification. For the case of R\"ossler oscillators, we note that there are also certain values of  coupling parameters
	and amplification where the synchronization doesn't exist or the synchronization
	can exist but without amplification. Using other
	systems with different topologies, we obtain some interesting results
	such as chimera state with amplification, cluster state with amplification and
	complete synchronization with amplification.
\end{abstract}

\maketitle

\begin{quotation}
	Research on multilayer networks has attracted a lot of
	attention in recent years in many areas of physics, engeneering, social sciences
	etc\cite{ref2,ref3,ref38,ref39,ref40,ref41}. Some emergent behaviors in such
	systems due to interaction among the dynamical units reveal a variety of
	interesting phenomena, such as synchronization \cite{ref1,ref2,ref4}, cluster
	formation \cite{reff11}, explosive synchronization \cite{ref12}, explosive
	desynchronization \cite{ref18}, chimera \cite{panaggio2015chimera,ref34,ref35,ref36,ref37}
	etc. Among these, synchronization and chimeras
	are the most widely studied. The notion of amplification is very important in science and technology. This work presents an investigation
	of different phenomena such as complete synchronization, cluster formation, phase
	synchronization and chimera states in a network with amplification. For an
	extended study we present three cases with three different topologies.
\end{quotation}

\section{Introduction}

The structure of many real world problems in nature, engineering, science and
technology is defined as a set of entities interacting with each other in
complicated patterns that can produce multiple types of relationships which
change in time and  exhibits a plethora of emergent patterns or behaviors as
synchronization, chimeras, chaos, consensus, cooperation, only to mention a few.
One very important ingredient behind most of those phenomena is the way in which
the particles or agents forming the systems interact, i.e., the topology of the
underlying network.

Network theory is used as an important tool for the modelling of dynamical
processes in complex systems \cite{ref1,ref2,ref3}. It plays also a major role
in the investigation of collective behavior. It finds many applications in
epidemiology where it is used to investigate epidemic spreading, in industry
where it is used in control
of behavior of machines, in dynamics of populations with the control of
displacement of the individuals, the cars, the drones or the airplanes.
According to the these applications, we can mention that the main objective is the
controllability of the network to lead to a certain state (it can be
synchronization, cluster state, phase-flip, chimera state, etc.) \cite{ref4,ref5,ref6,ref8,ref34,ref35}. Thus, the investigation of
the dynamics of the networks need the expertise of some mathematical tools such
as the Master Stability Function (MSF) developed by Pecora and Carroll,
the transversal Lyapunov exponent, the correlation between the oscillators of
the same or of a different layer, etc. One of the best methods to study the
stability of the synchronization in the network is the Master Stability Function
\cite{ref9}. This method is used for coupled identical oscillators.

Many researchers are studying  several phenomena which take
place in  multiplex networks. Such interest is motivated in understanding how the complete or partial synchronization
occurs in this type of systems and also because the topologies of multiplex networks
appear in several natural and tecnological systems.
Multiplex network  may be described as being a collection of two or more coupled
networks where a set of networks is connected by links where the interactions are of different types \cite{boccaletti2014,ref42}. These links characterize the connections
existing between any node or network of the multiplex network. Many recent works addressing multilayer structures
and systems were summarized in \cite{boccaletti2016}.

The study of inter-layer synchronization in non-identical multilayer networks
was address in \cite{leyva2017}. The authors were able to show an analytical
treatment for a
two-layer multiplex using the Master Stability Function method. One interesting
outcome was to predict the effect that missing links in one of the layers has on
the inter-layer
synchronization. Later,  in \cite{gosh2018} it was found that  a sparse
inhomogeneous second layer can promote chimera states in a sparse homogeneous
first layer. The study of synchronization of non-identical multilayers is very
recent, thus many collective behavior properties and patterns may unravel.

Here, we consider a network with two layers where we choose an
all to all coupling in the layer for the case of R\"ossler oscillators, a
ring with bidirectional coupling in each layer for the case of jerk oscillators
and aglobal mean field coupling for the last case mentioned above. The connection between the systems of 
both layers (interlayer coupling) is unidirectional. The main goal of this work is to investigate in each case the dynamics
of each layer as well as the whole network with amplification in the second layer. We found the key values of the parameters to control synchronization with and without
amplification.

The remainder of this work is organized as follows. In section II, we present
our multilayer network with the mathematical description of the model and the
systems. The dynamics of the main case is presented in section
III emphasizing on the intralayer and interlayer coupling in which numerical simulations
are done. An application to another two systems has been studied
in section IV and finally, we present the conclusions in section V.

\section{Multilayer network}

\begin{figure}[htp]
	\begin{minipage}[b]{8cm}
		\begin{center}
			\includegraphics[width=5cm, height=4cm]{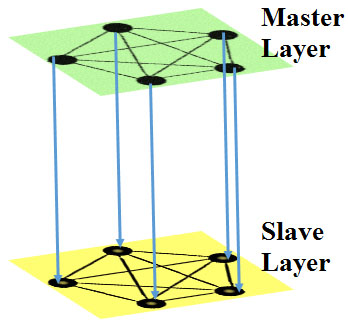}
			\caption{\footnotesize{Schematic representation of a
					network with two layers of interaction.}}
			\label{Model}
		\end{center}
	\end{minipage}
\end{figure}

The model consists in a multilayer network constituted of $N$ nodes connected in each layer, which can be 
represented by a $2N \times 2N$ adjacency matrix $A_{ij}$ where the elements of this matrix are respectively $1$ if 
the nodes $i$ and $j$ are connected and $0$ if not. Based on Ref.\cite{gosh2018}, the adjacency matrix of the whole 
network consisting two layers can be expressed as follow:
	\begin{align}\label{adjm}
	\textbf{A} = \left( {\begin{array}{*{4}{c}}
		A^{1} & 0 \\
		I & A^{2} \\
		\end{array}} \right)
	\end{align}\\
	where $A^{1}$ and $A^{2}$ are the $N \times N$ adjacency matrix modelling the intralayer connectivity in 
	the first and second layer respectively. $I$ is an $N \times N $ identity matrix representing
	the unidirectionnal interactions (Layer 1 -> Layer 2) between the oscillators with the same index in both layers. 
	The use of the 
	null matrix is justified by the non-existence of a connection from the slave layer to the master layer. 
	In the following, we consider a model of multilayer network constituted of $N$ nodes in each layer connected using an all-to-all coupling scheme in each layer (see Fig.\ref{Model}). To each node corresponds a nonlinear autonomous R\"ossler oscillator as described in\cite{rossler1979continuous}. Notice that it 
	is this combined oscillator which defines our network as a two-layer system made up of a driving system and a slave one.

The dynamics of the first layer also considered as the driver for the network is described 
by Eq.\ref{layer1} where $\epsilon_1$ is the intra-layer coupling strength for the first layer. \\

\begin{equation}
\label{layer1}
\begin{array}{cc}
\left\{ \begin{array}{l}
\dot x_{i}^{1} =  - x_{i}^{2} - x_{i}^{3} +
{\epsilon_{1}}\sum\limits_{k = 1}^N A_{ik}^{1} (x_{k}^{1}
-x_{i}^{1}),\\
\dot x_{i}^{2} = x_{i}^{1} + ax_{i}^{2},\\
\dot x_{i}^{3} = bx_{i}^{1} + x_{i}^{3}(x_{i}^{1} - c).
\end{array} \right.

\end{array}
\end{equation}\\

The second layer has exactly the same intra-layer topology with similar systems in the nodes. 
The choice of systems 
in the second layer follows Louodop et al. \cite{ref29} where the authors show in Appendix A that, this 
form of coupling 
between elements of different layers produces generalized synchronization\cite{ref29}. 
Notice that it is  
this system defined in Louodop et al. \cite{ref29}, what defines which oscillators in Layer 2 interact with those in 
Layer one, while keeping 
the same all to all intra-layer topology. 
Therefore, the dynamics of the second 
or slave layer is given by Eq.\ref{layer2}:

\begin{equation}
\label{layer2}
\begin{array}{cc}
\left\{ \begin{array}{l}
{{\dot y}_{j}^{1}} =  - {y_{j}^{2}} - {y_{j}^{3}} + {\epsilon_2}\sum\limits_{k = 1}^N A_{jk}^{2}
({y_{k}^{1}} - {y_{j}^{1}}) + C_{0}(x_{j}^{1} - C_{2}y_{j}^{1}),  \\
{{\dot y}_{j}^{2}} = \frac{{x_{j}^{1}} + a{x_{j}^{2}}}{C_2},\\
{{\dot y}_{j}^{3}} = b{y_{j}^{1}} +
\frac{{{x_{j}^{3}}{x_{j}^{1}}}}{{{C_2}}} - c{y_{j}^{3}}.
\end{array} \right.
\end{array}
\end{equation}

Here, $\epsilon_{2}$ is the intra-layer coupling strength and $C_0$ is the interlayer coupling strength. 
It is important to mention 
that, this inter-layer coupling exists only between the oscillators with the same index (i.e. for $i=j$ where the $j$ index 
runs from $1$ to $N$).
$C_{2}$ is the parameter of proportionality named amplification coefficient. This term or interaction via a $conjugate$ variable 
has been used in the literature to model $revival$ \cite{zou2015restoration, ghosh2015revival} 
as well as $amplitude$ $death$ \cite{karnatak2007amplitude}.
For all these layers we consider $a=0.36$, $b=0.4$ and  $c=4.5$ and we note that at these values of the 
parameters the systems operate in the chaotic regime\cite{rossler1979continuous}. This 
topology of connectivity between the master and the slave layer imposes generalized 
synchronization between both layers in the absence of intralayer coupling because as it is 
conceived (see Appendix A), the slave layer is supposed to function as an observer of the 
master layer with certain conditions $C_2 \neq 0$. 
In this work, we were interested in the 
notion of synchronization with amplification (or reduction) depending on the value of $C_2$. 
For $C_2 > 1$ we have an amplification of the systems of the master layer (or a reduction of the systems of the 
slave layer) and conversely for $C_2 < 1$. It should be noted that when $C_2$ takes negative values there is an 
anti-synchronization between the systems of the master layer and those of the slave layer having the same index. 
The amplification coefficient must be different from zero ($C_2\neq0$) and bounded, because if $C_2=0$
the systems of the 
slave layer will diverge and if $C_2$ is too large the slaving factor will tend to zero. 
Therefore we keep $C_2$ in the interval between $0.005$ and $2$. This type of topology finds 
applications in many domains such as aircraft control where recently they have proven that an optimal control permitted 
to regulate the air traffic in the sky \textcolor{blue}{\cite{cafieri2018combination}}. Considering the domains of application of this network topology, 
it seems rather important to investigate the dynamics of this network with different parameters, which is presented in the next section.

\section{Dynamics of networks}

	\subsection{Dynamics of Different Layers}
	
	Considering the topology given in Fig.\ref{Model} and the mathematical equations
	for both layers  Eq.\ref{layer1} and
	Eq.\ref{layer2}, respectively, we are going to investigate
	numerically the dynamics in the different layers using the MSF described in Appendix B. 
	A numerical calculation is done using Runge-Kutta fourth order
	for a long time simulation and the permanent solutions are considered at
	$t_{min}=0.6$ $ t_{max}$.
	
	\begin{figure*}[htp]
		\begin{center}
		\includegraphics[width=5.5cm, height=4cm]{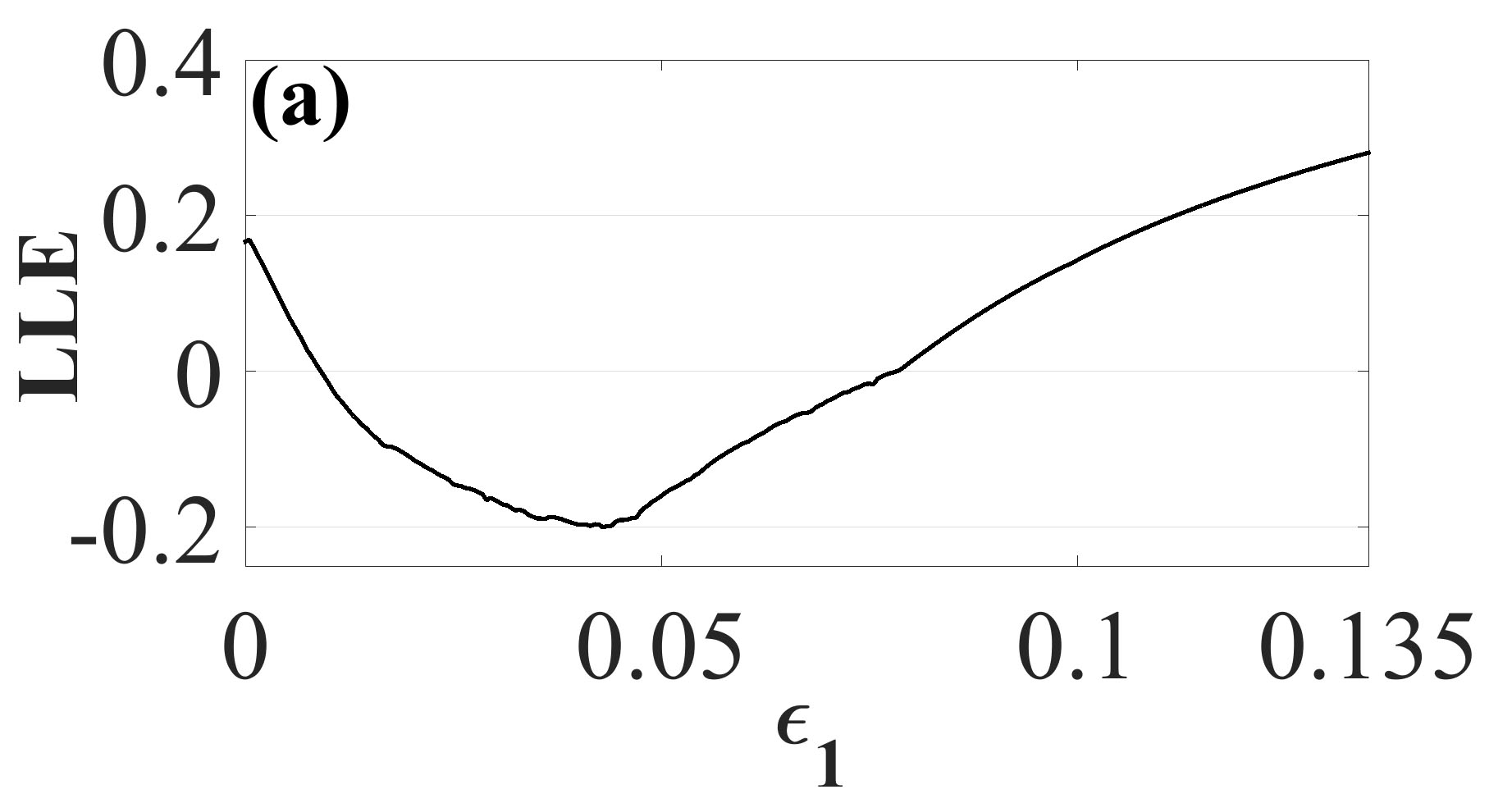}
		\includegraphics[width=5.5cm, height=4cm]{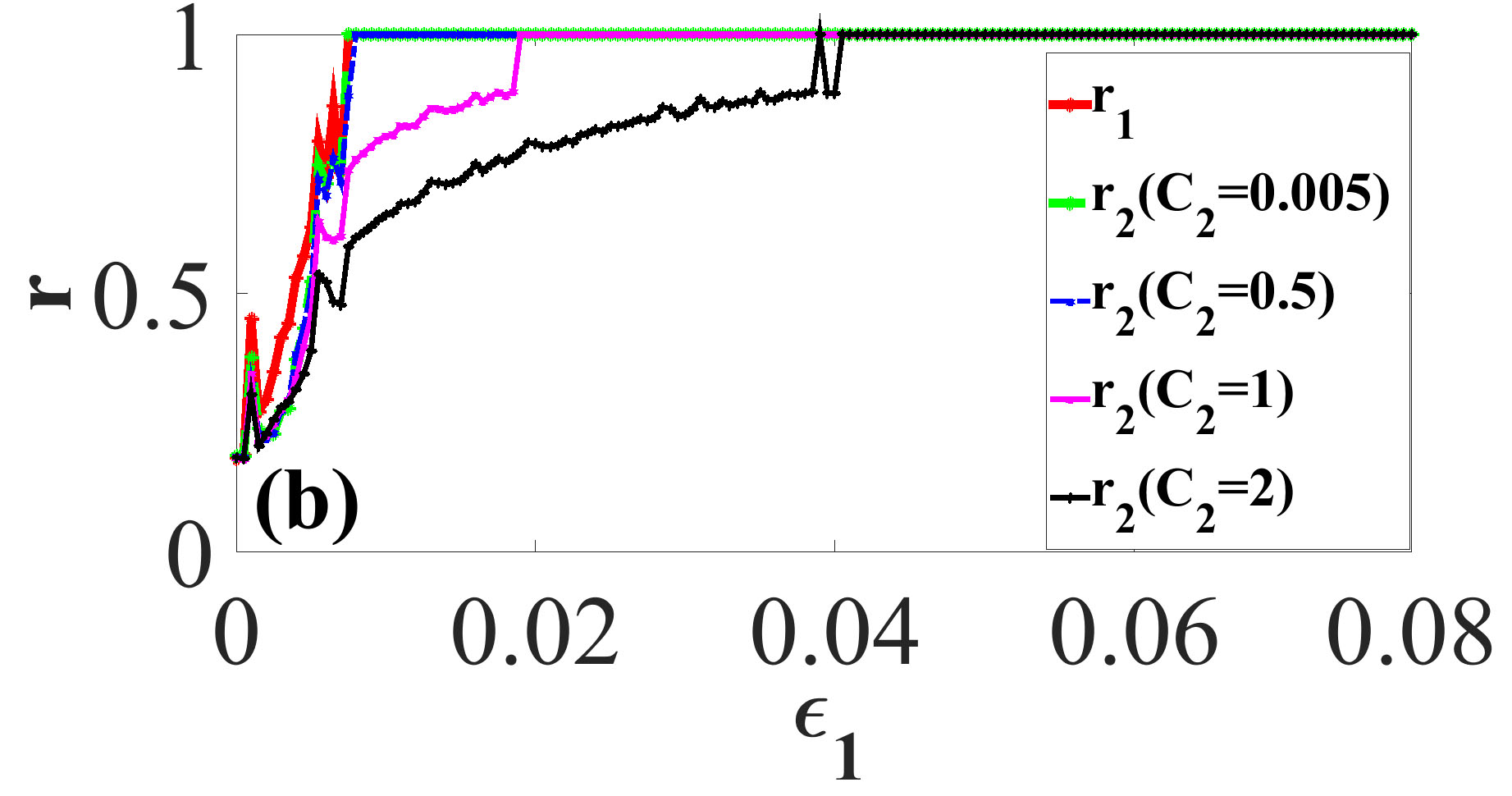}
		\includegraphics[width=5.5cm, height=4cm]{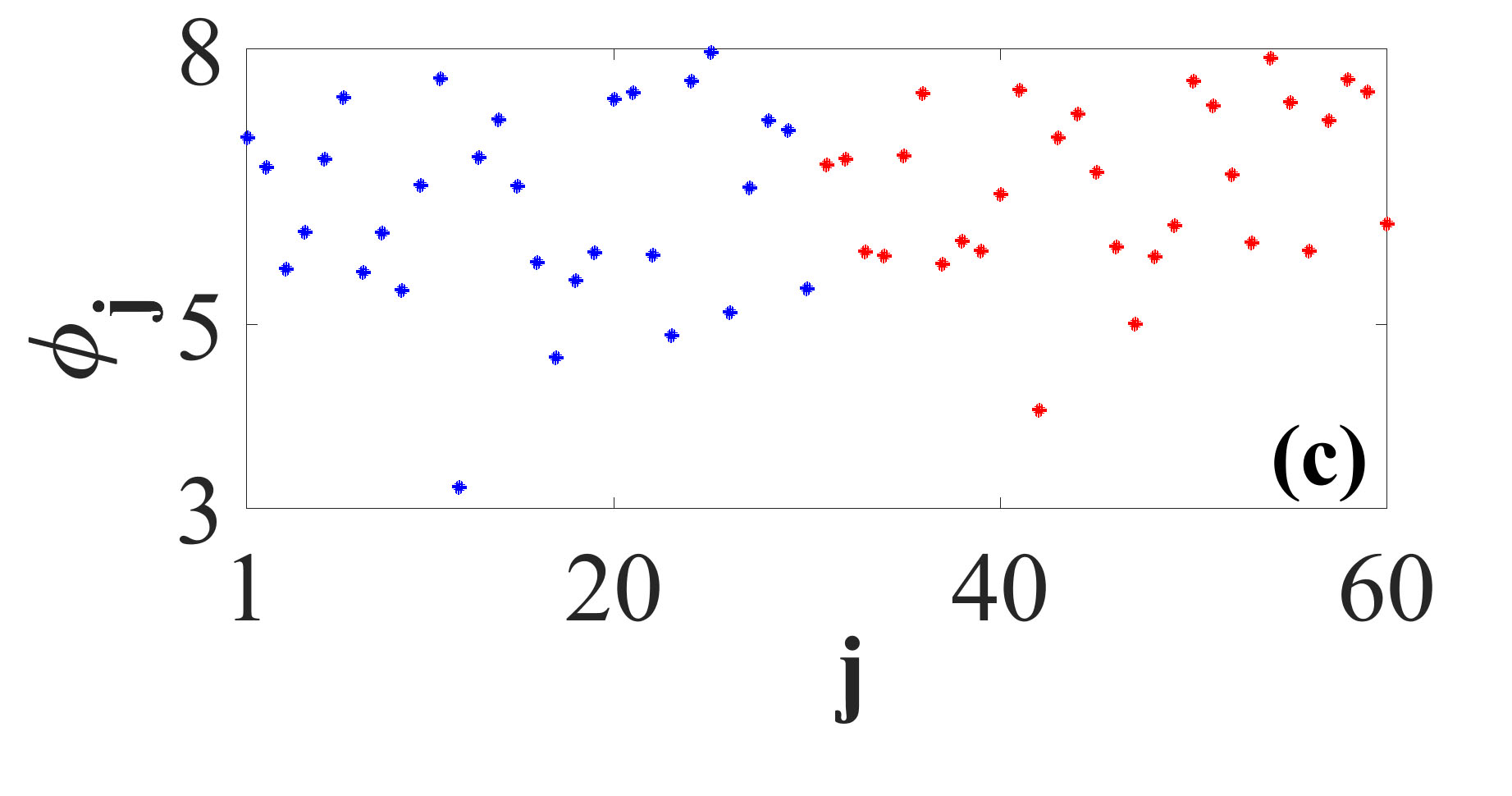}
		\includegraphics[width=5.5cm, height=4cm]{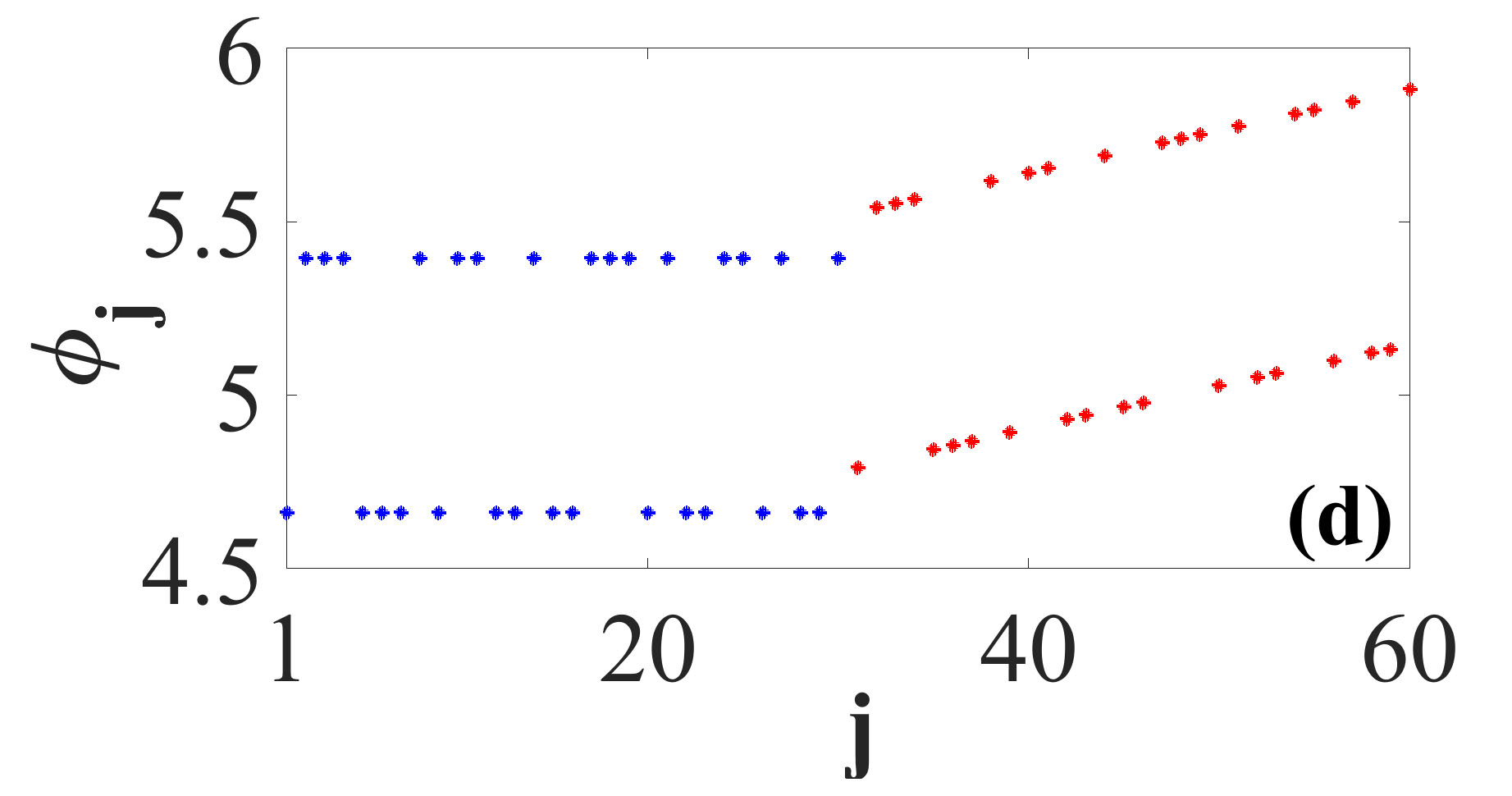}
		\includegraphics[width=5.5cm, height=4cm]{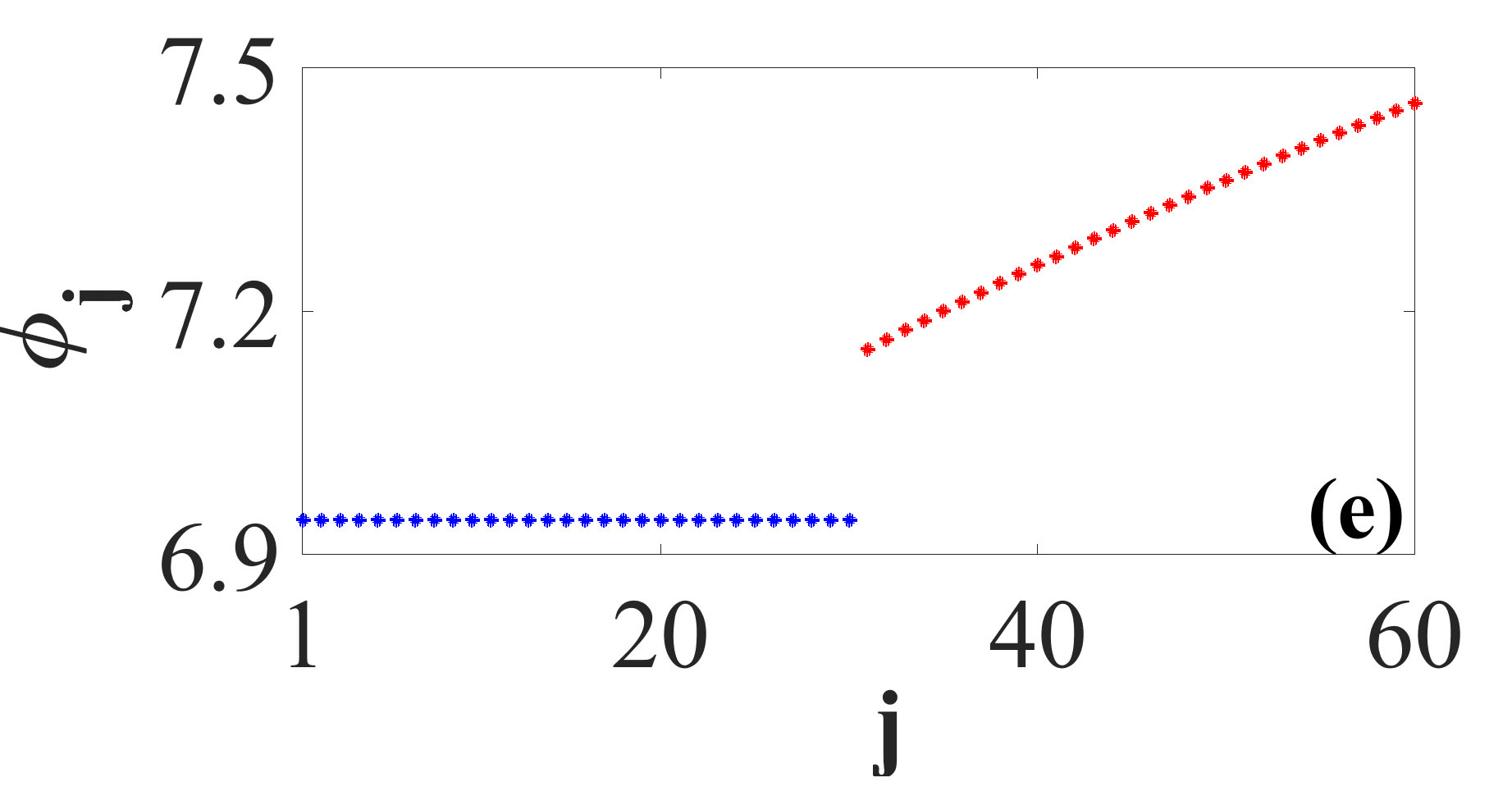}
		\includegraphics[width=5.5cm, height=4cm]{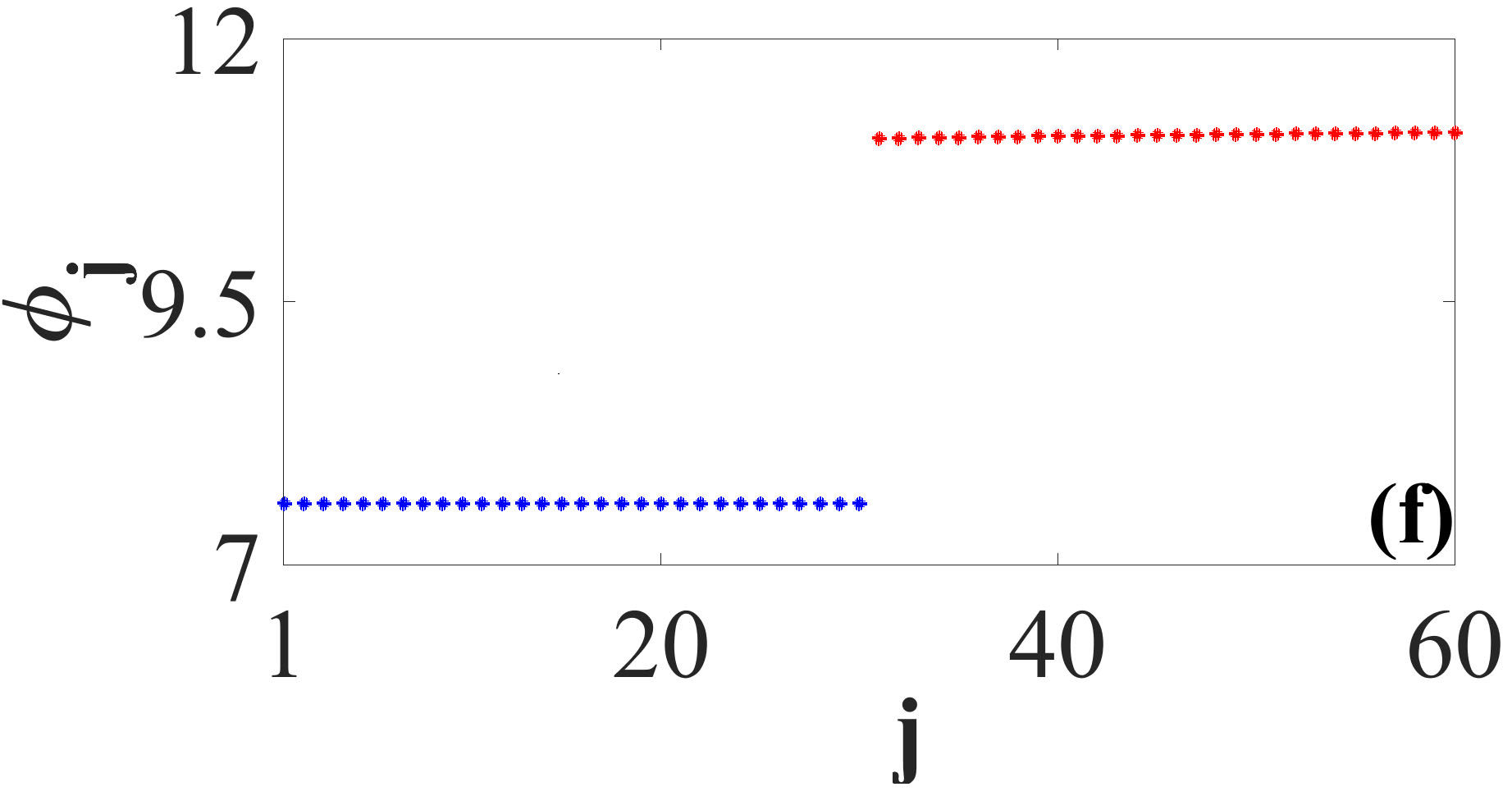}
		\includegraphics[width=8.5cm, height=6cm]{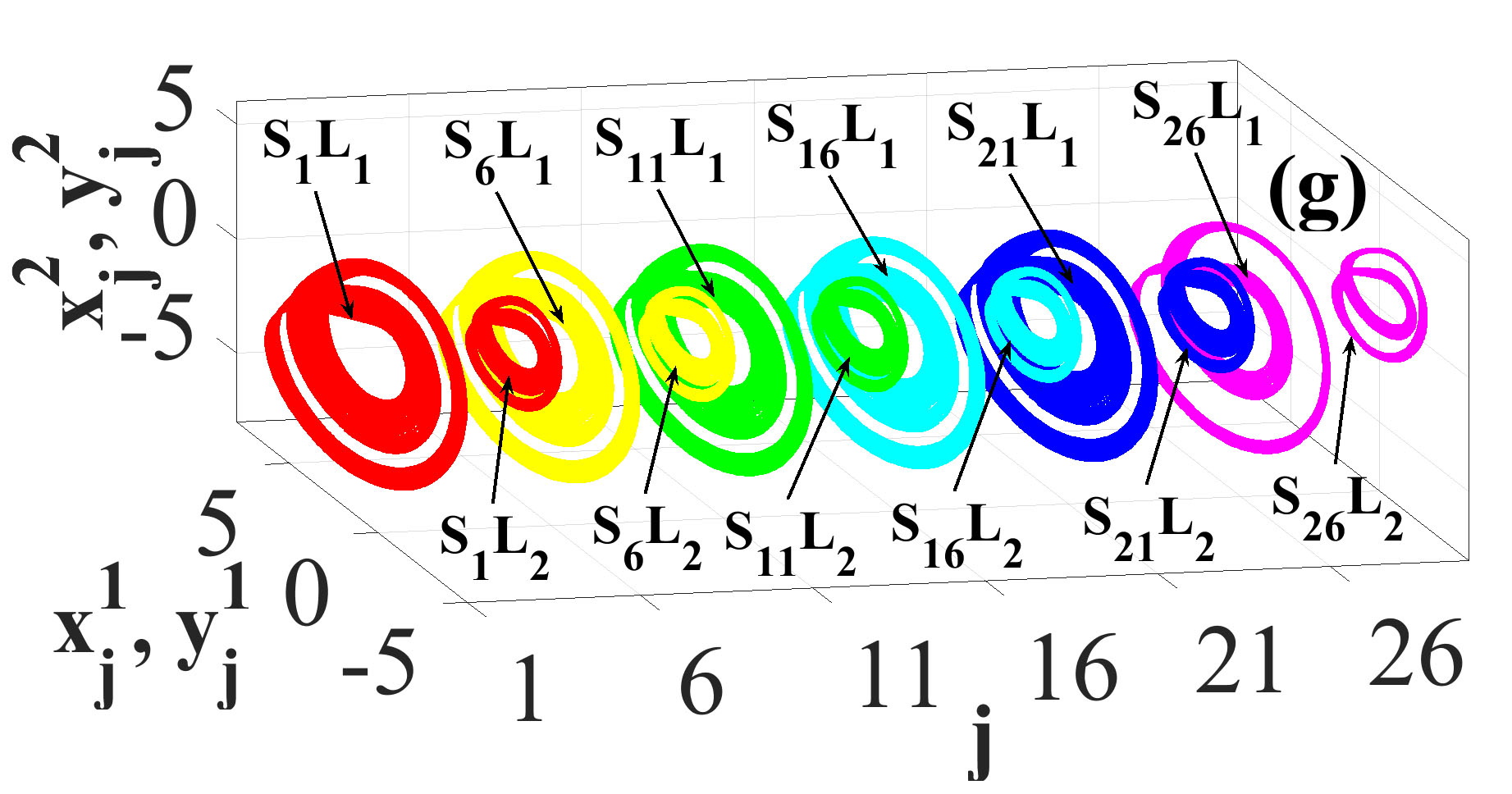}
		\includegraphics[width=8.5cm, height=6cm]{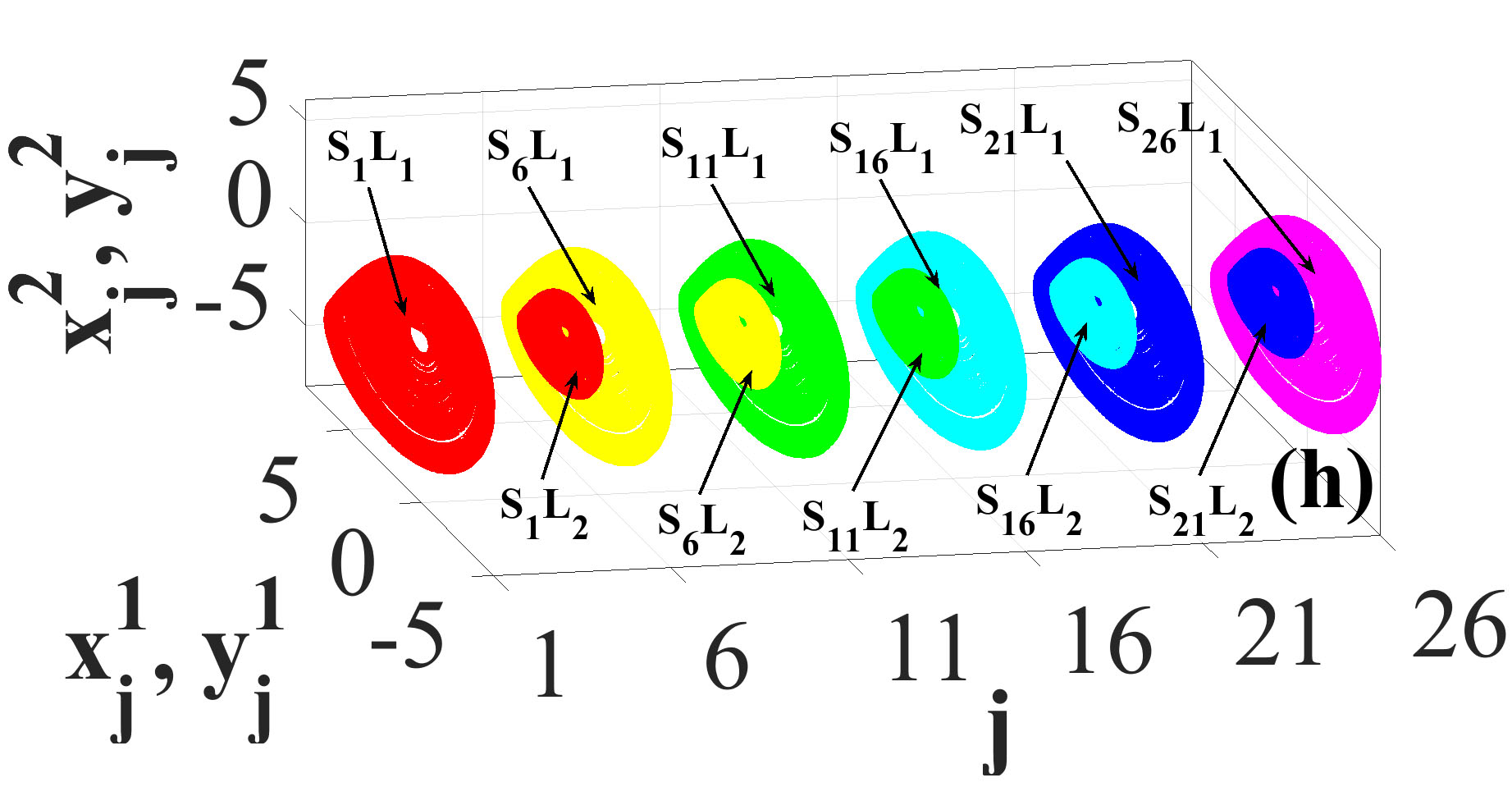}
     	\end{center}
		\caption{\footnotesize{Dynamics of the two layers of the network: (a)
				Master stability function of the first layer as a function of the intralayer coupling considering 
				$\epsilon_2=10\epsilon_1, C_0=1$ and $C_2=2$. (b) Order parameter showing the dynamics of the slave layer 
				for different values of the	amplification coefficient for $\epsilon_2=10\epsilon_1$ and $C_0=1$.
				(c,d,e and f) Mean phase of the first and second layer respectively for	$\epsilon_1=0.005, 0.007, 0.009, 0.04, C_0=1$ 
				and $C_2=2$. (g)  Attractors of the oscillators $1$, $6$, $11$, $16$, $21$ and $26$ in the master and slave layer for $\epsilon_1=0.007$.
				(h) Attractors of the oscillators $1$, $6$, $11$, $16$, $21$ and $26$ in the master and slave layer $\epsilon_1=0.009$ respectively. 
				$S_iL_l$ means system $i$ of the layer $l$ ($i=1,2,...,N$ and $l=1,2$).}} 	
		\label{MSF1}
	\end{figure*}
	
	To illustrate the behaviour of the oscillators of the first layer we present in
	Fig.\ref{MSF1}(a) the Master Stability Function or Largest Lyapunov Exponent ($LLE$) of the 
	variational equation Eq. \ref{VE2}, used to characterize the stability of the synchronization in the first layer. 
	In this figure there are two important regions in terms of characterization of the stability of the synchronization: 
	if the $LLE\leq0$ the synchronization is stable and the $LLE>0$ the synchronization
	is unstable. By varying smoothly the intralayer coupling	$\epsilon_{1}$,
	this figure shows that, when $\epsilon_{1}$ increases the systems evolve to the synchronous state at $\epsilon_1=0.009$. 
	This synchronization is obtained at a minimal value of the coupling strength in the first layer and we note that the
	synchronization in the slave layer is highly influenced not only by the
	coupling but also by the amplification, this can be seen on the behavior of the
	order parameter \cite{chen2017order,hong2000phase}  which is plotted in Fig.\ref{MSF1}(b) as a 
	function of $\epsilon_1$. The procedure to compute the order parameter is explained in Appendix C. 
	It should be noted that the notion of amplification here is related to the amplitude of the oscillations of the state 
	variables of the systems. We are talking about synchronization with amplification if and only if the ratio $X/Y=C_2$ is 
	respected, as shown
	in Appendix A. 
	So, we investigate here the impact of the amplification parameter on the dynamics of the
	network, where the order parameter of the slave layer is plotted for different
	values of $C_2$. According to this figure although interaction between the
	layers does not impede synchronization in the slave layer, it becomes more
	effective for small values of $C_2$. Based on the demonstration given in Appendix A synchronization occurs 
	at $Y=\frac{X}{C_{2}}$. Therefore, when $C_2<1$ we obtain an amplification in the slave layer and respectively a reduction 
	in master layer and vice-versa when $C_2>1$. Thus, if we need to achieve synchronization in both layers at the same 
	value of the interlayer coupling, $C_2$ must be very small, leading to a significant amplification at the second layer. 
	Also from Fig.\ref{MSF1}(b) we see that for the minimum value of $C_2$ = $0.005$, the synchronization in both layers 
	happens at the corresponding values of $\epsilon_1$ and $\epsilon_2$ ($\epsilon_2=10\epsilon_1$), this is represented 
	in Fig.\ref{MSF1}(b), where the order parameter $r_1$ and $r_2$ for both layers is seen to reach the 
	value for complete synchronization at the same point. This value of $C_2$ produces an amplification of around $200$ of 
	all the variables of the master layer $  (Y \approx 200X)$ in the slave layer. In Fig.\ref{MSF1}(c) we plot
	the mean phase of the driver and the slave layer (the first 30 systems are for the
	driver layer and the rest for the slave layer) for the value $\epsilon_1=0.005$
	of the intralayer coupling. The situation shown here is confirmed by the
	Fig.\ref{MSF1}(a) ($LLE>0$). If we consider $\epsilon_1=0.007$,
	Fig.\ref{MSF1}(a) shows that the Largest Lyapunov Exponent is non negative but
	very close to zero, then in the Fig.\ref{MSF1}(d) we show the mean phase
	where the
	master layer has a two cluster synchronization with equal phases
	\cite{reff11,refff11,ref12,reff12,ref13}.  In the slave layer, while the clusters follow the systems 
	with the same index as that of the master layer, we see an oblique sliding of the
	systems reminding of a splay state.  To better appreciate the dynamics of the oscillators in this behaviour, we show in 
	Fig.\ref{MSF1}(g) the attractors of the oscillators labeled $1, 6, 11, 16, 21, 26$ in the master and slave layers
	for  $C_2=2$ and $\epsilon_1=0.007$ ( $LLE\ge0$). 
	 For $\epsilon_1=0.009$ (with $LLE<0$), we obtain
	Fig.\ref{MSF1}(e)
	which represents the synchronization in the first layer and a coherent oblique
	sliding in the second layer. By computing the phase difference between
	consecutive oscillators (here consecutive refers to the indices of the
	oscillators) we verified that the phase distance between oscillators of
	consecutive
	index
	in the slave layer is constant, therefore the second layer presents indeed a
	phenomenon
	of splay \cite{ref14,ref15}. In Fig.\ref{MSF1}(f) we show for $\epsilon_1=0.04$
	the phase synchronization in both layers but not at the same value of the mean
	phase. So, for the multilayer network, the dynamics is equivalent to that of two
	clusters. To illustrate the dynamics of the system
		we show in Fig.\ref{MSF1}(h) the attractors of some
		oscillators (labeled $1, 6, 11, 16, 21, 26$) for $\epsilon_1=0.04$ to
		appreciate the behaviour of the entire
		network.
	
	Therefore, we can conclude that, according to the different values of the
	intralayer coupling the network leads to different phenomena such as, cluster
	synchronization, splay, synchronization and the stability of this
	synchronization is confirmed by the MSF.

	\subsection{Impact of the interlayer coupling and amplification on the dynamics of the network}
	
	In the previous Sections we have shown the influence of the intralayer coupling on
	the dynamics of the network. We see that
	the network presents many
	phenomena depending on the amplification coefficient $C_2$ and the interlayer
	coupling $C_0$. In this section, our goal is twofold: first we investigate the
	behaviour of the network under the impact of these two parameters and then we
	show the effect of the amplification as well as  its effectiveness in the
	network. We keep the intra-coupling constants ($\epsilon_1, \epsilon_2$) fixed,
	varying smoothly the amplification coefficient ($C_2$) from 0.005 to 2 and the
	inter-coupling ($C_0$) from 0 to 20.
	\begin{figure}[htp!]
		\includegraphics[width=\linewidth]{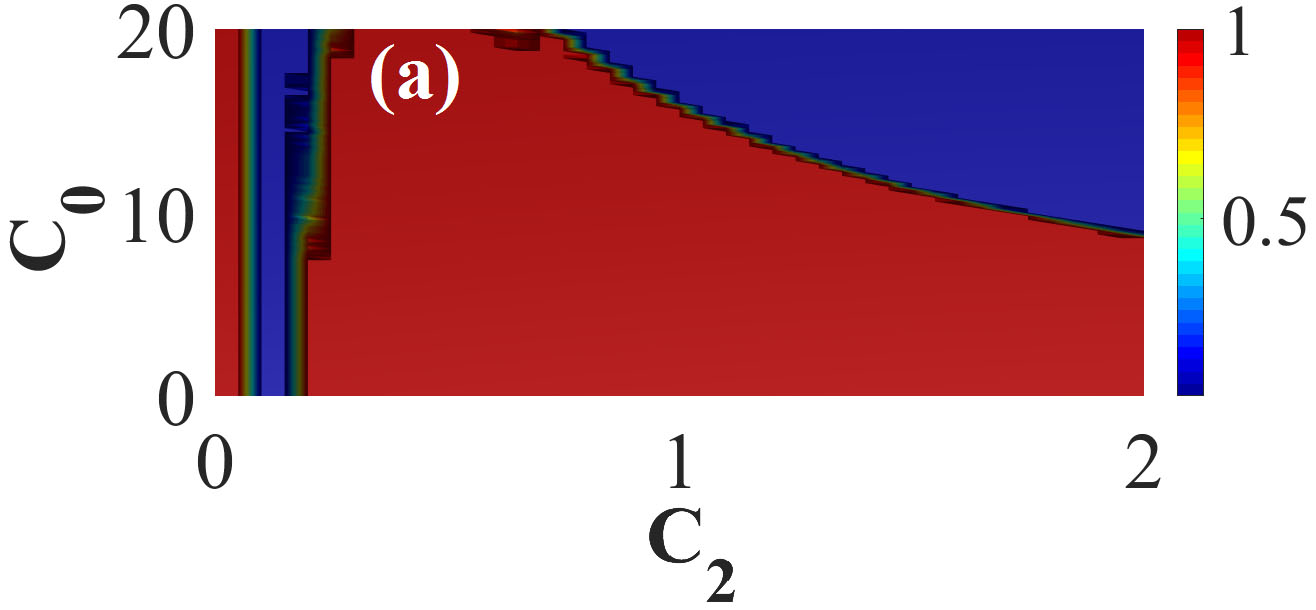}
		\includegraphics[width=\linewidth]{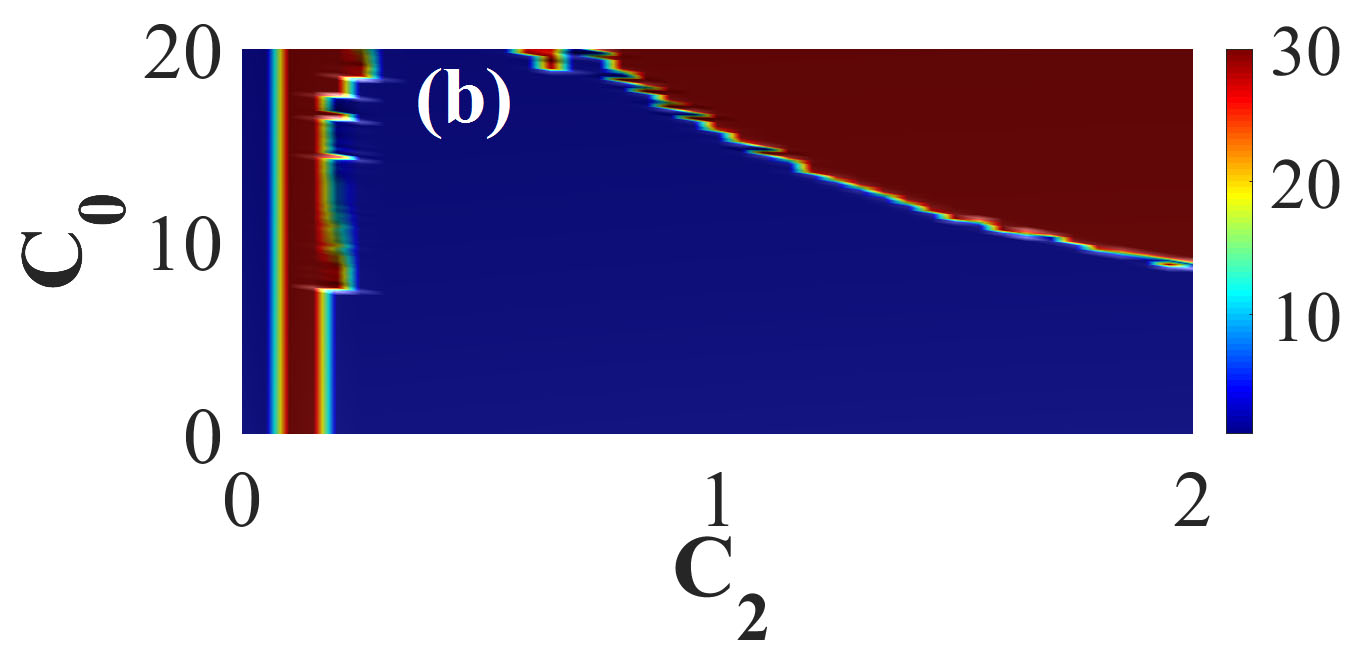}	
		\includegraphics[width=\linewidth]{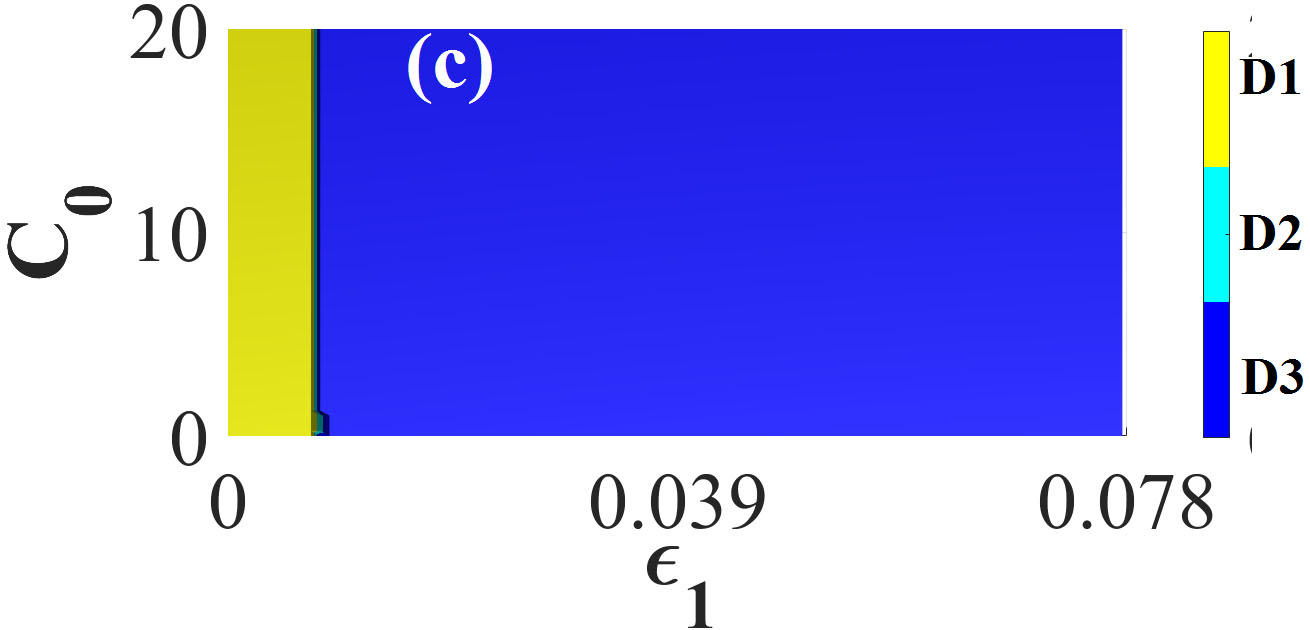}
		\includegraphics[width=\linewidth]{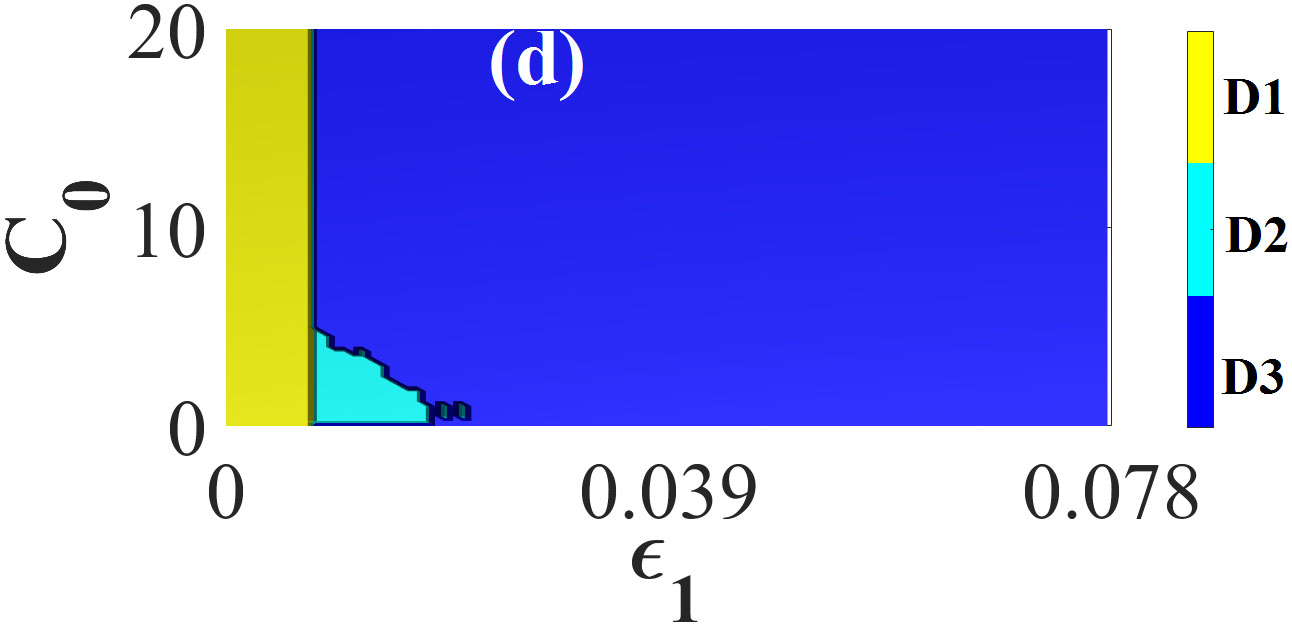}
		\includegraphics[width=\linewidth]{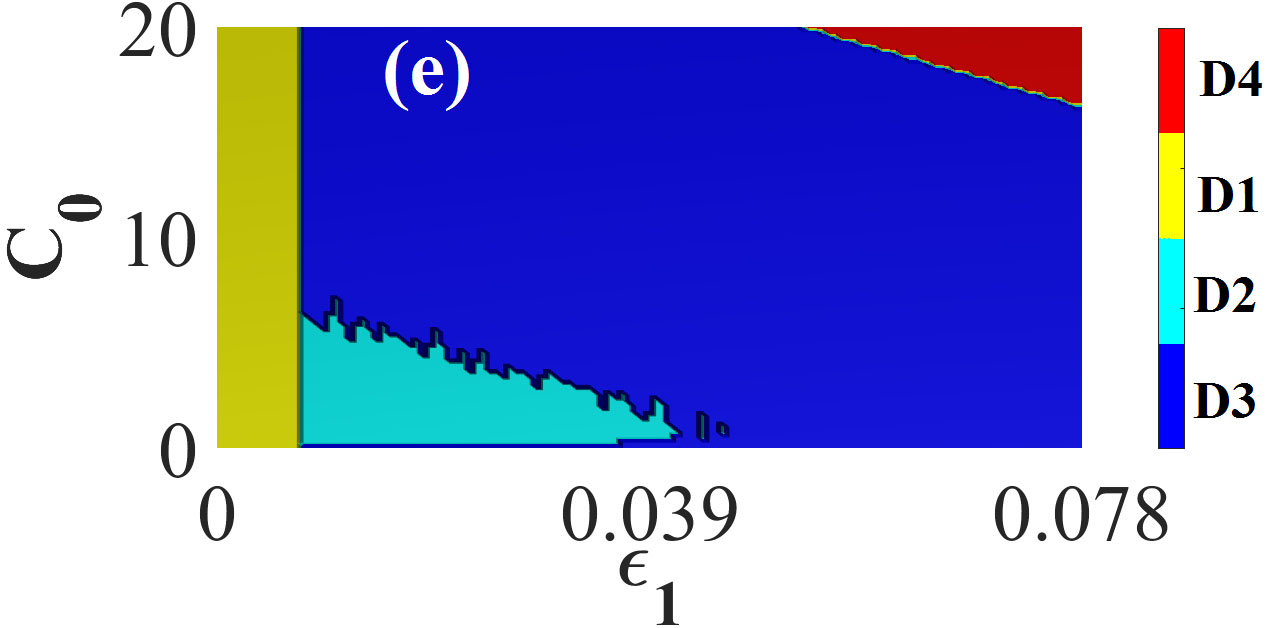}
		\caption{\footnotesize{Network synchronization regions: (a) Order parameter of the multilayer network for
				$\epsilon_1=0.03$ and $\epsilon_2=0.25$ as a function $C_0$ and $C_2$. (b) Number of states in the slave layer for $\epsilon_1=0.03$ 
				and $\epsilon_2=0.25$ as function $C_0$ and $C_2$. Two parameter phase diagram by simultaneously varying the intra-layer 
				coupling $\epsilon_1$ and inter-layer coupling $C_0$ with
				$\epsilon_2=10\epsilon_1$ and for different values of the
				amplification
				coefficient $C_2$: (c) $C_2=0.5$, (d) $C_2=1$  and (e)
				$C_2=2$. }}
		\label{DL1}
	\end{figure}

\begin{figure*}[htp]
		
		\includegraphics[width=9cm, height=4cm]{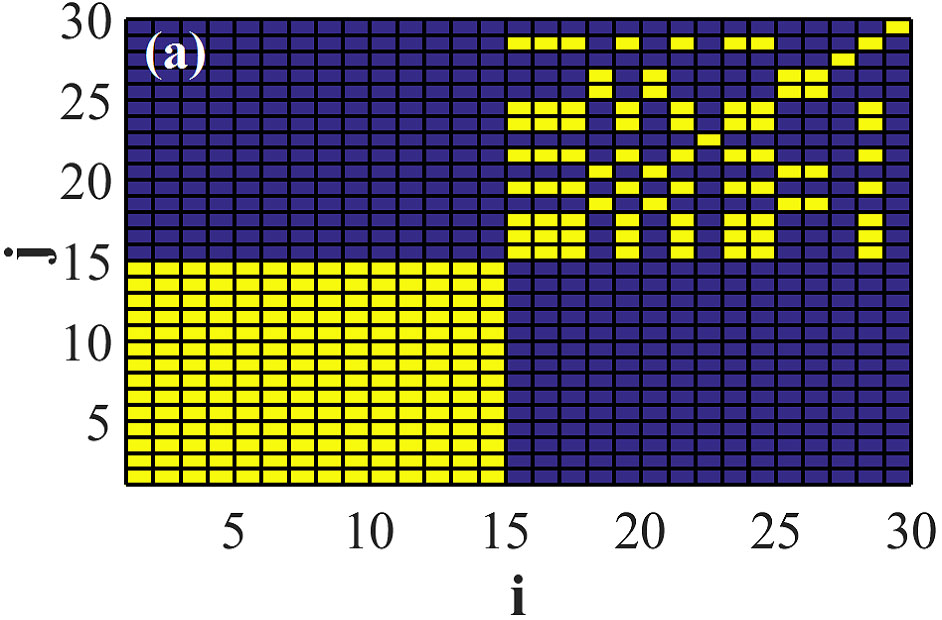}
		\includegraphics[width=8cm, height=4cm]{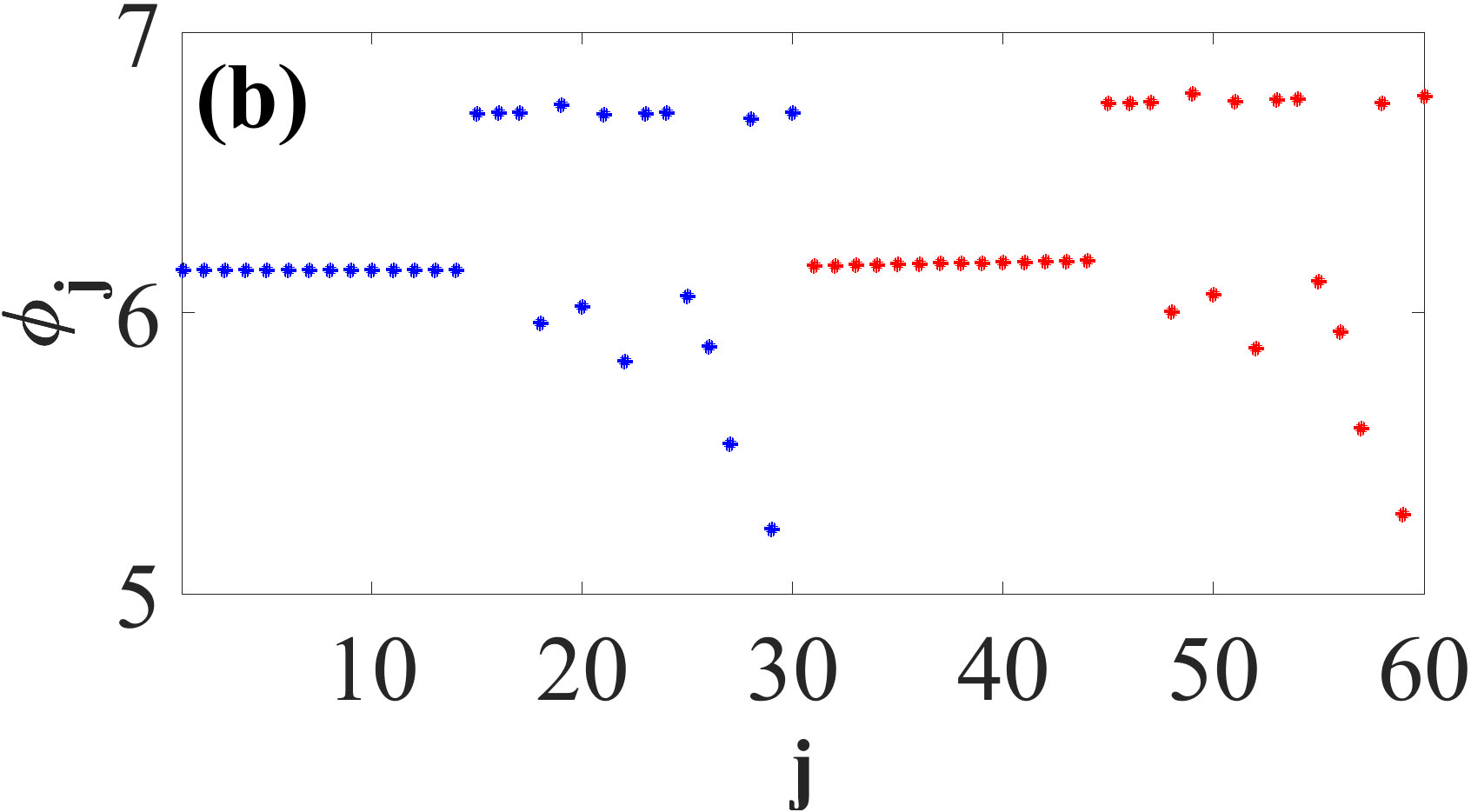}
		\includegraphics[width=17.0cm, height=7.5cm]{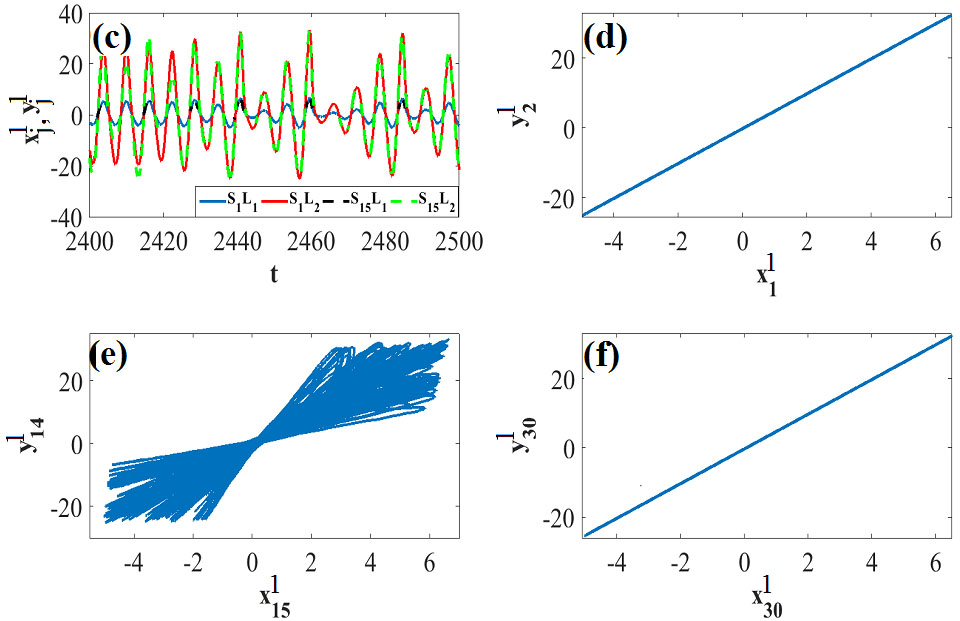}
		\includegraphics[width=9cm, height=4cm]{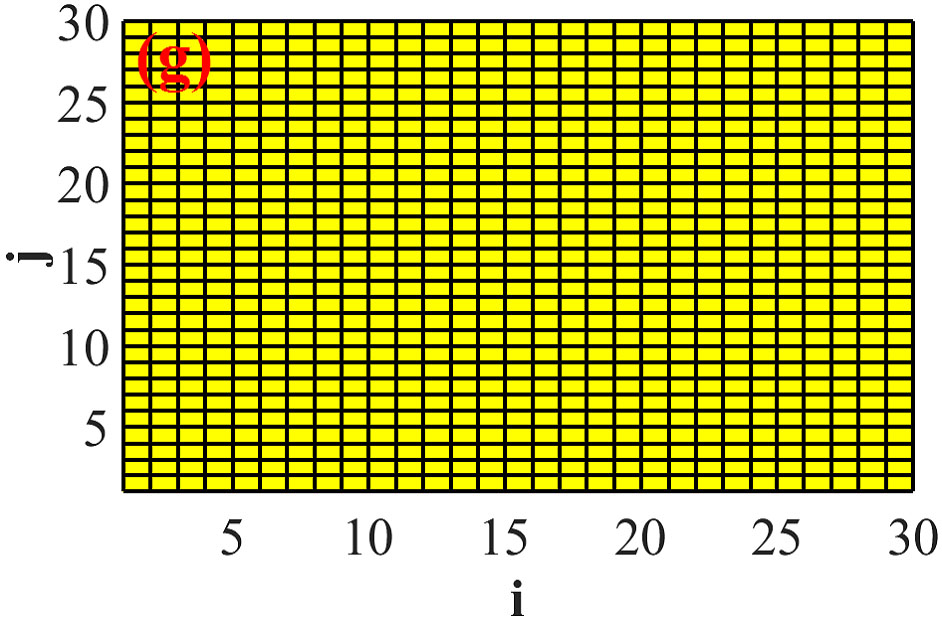}
		\includegraphics[width=8cm, height=4cm]{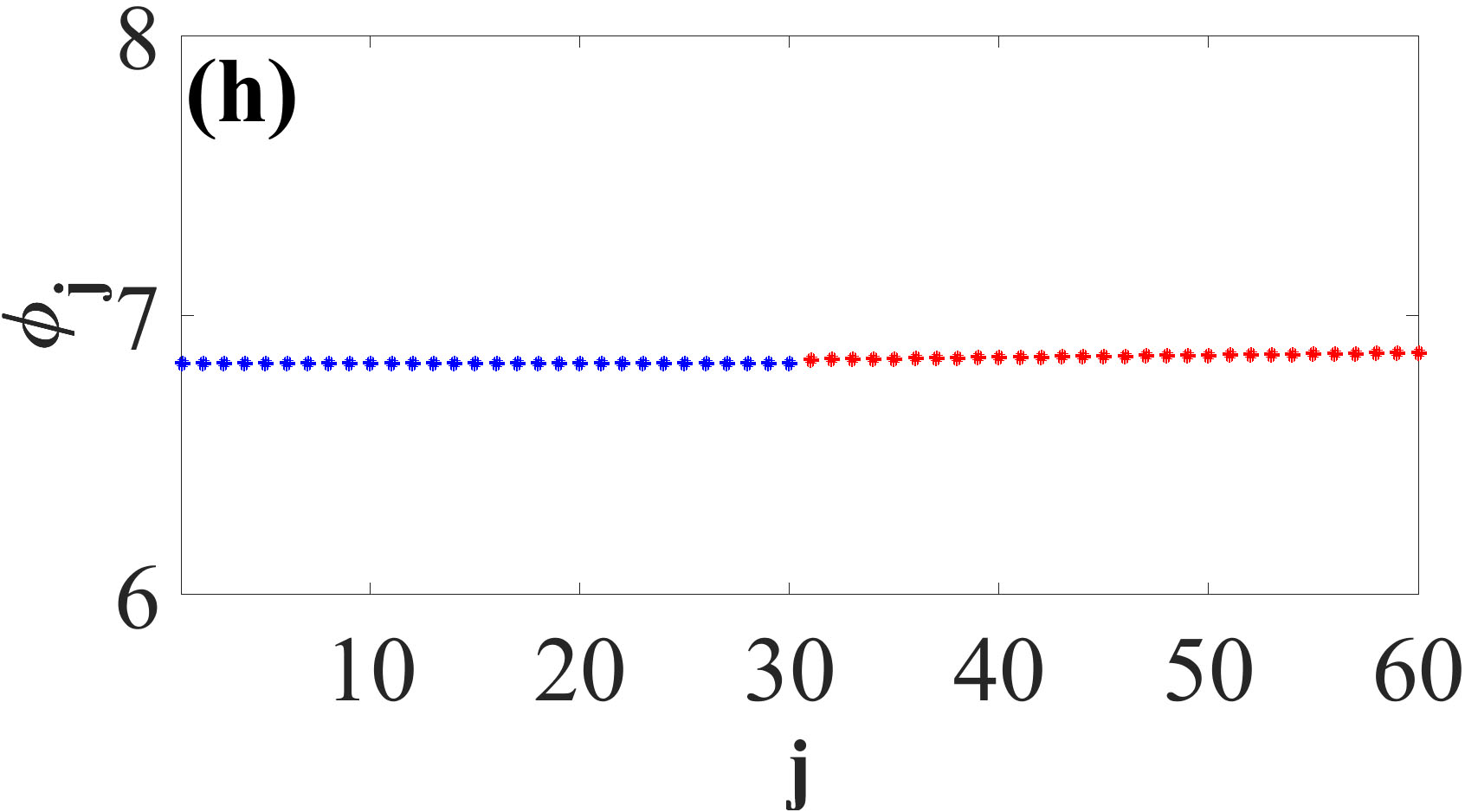}
		\includegraphics[width=17.0cm, height=7.5cm]{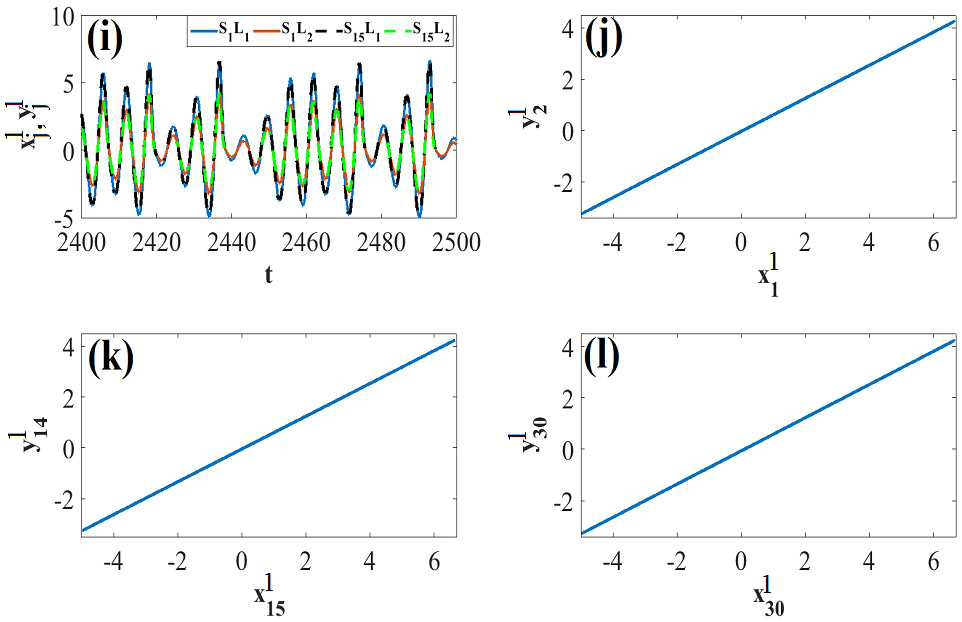}
		
		\caption{\footnotesize{Dynamics of the multi-layer network for
				$C_0$ and $C_2$: (a) Correlation between $x_{i}^{1}$ and $y_{j}^{1}$ (Yellow color 
				indicates the oscillators in the synchronized state  and blue the oscillators in an 
				unsynchronized state.) (b) Mean phase
					and (c) time series of some oscillators of the network for $C_2=0.2$ and $C_0=10$. 
					$S_i L_l$ means system $i$ of 
					the layer $l$ (i = 1, 2, ..., N and l = 1, 2).
					(d,e,f) Synchronization between some oscillators in the first and second layer
					of the network for $C_2=0.2$ and $C_0=10$. (g) Correlation between $x_{i}^{1}$
					and $y_{j}^{1}$. (h) Mean phase and (i) time series of some oscillators of the network for
					$C_2=1.55$ and $C_0=10$.  (j,k,l) Synchronization between some oscillators in the
					first and second layer of the network for $C_2=1.55$ and $C_0=10$.
					 Figures (d,e,f) and (j,k,l) are special cases of the correlations presented in Figures 4(b and g). However, these figures even if they may confuse
					for being similar show that amplification clearly depends on the value of $C_2$, and also show that in some cases the synchronisation of oscillators of different index 
					is possible (see Figs. 4d and 4k).}}
		\label{DL}
	\end{figure*}

	As mentioned in Fig.\ref{MSF1}(b), the synchronization in the slave layer is
	imposed by the synchronization in the master layer. According to
	the
	literature \cite{ref16,ref17}, in order to bring our
	multilayer network
	towards a desired
	behaviour such as synchronization, cluster formation, splay and so on, it
	suffices to adjust the coupling.
	Although this is usually the case, in our system we have two important
	parameters acting as
	interlayer couplings ($C_0$ and $C_2$) with the
	difference that one of them ($C_2)$ allows us to increase or decrease the
	amplitude of the signal in one of these layers. To illustrate the evolution towards the synchronization as a function of the amplification parameter $C_2$ and the
	interlayer coupling $C_0$ we show in Fig.\ref{DL1}(a) the order parameter of the
	slave layer, since we are interested in synchronization throughout the network.
	By considering the intralayer coupling $\epsilon_1=0.03$ and $\epsilon_2=0.25$ all the oscillators of the first layer are synchronized but due to the effects of
	the amplification parameter $C_2$ and the interlayer coupling, the dynamics of
	the slave layer can not be the same. Therefore, the dynamics of the
	multilayer network  and particularly the slave layer is greatly affected by  $C_0$ and $C_2$. 
	This can be observed in Fig.\ref{DL1}(a) where we plot the order parameter 
	of the second layer. In red we represent the region where synchronization and amplification is obtained. We note
	that in  this red zone the relation $X=C_2Y$ is verified with a precision of
	$10^{-4}$, while in the blue domain it is possible to obtain synchronization in
	the first layer only but not in both layers of the network.
	This information is corroborated by the number of states of the second layer 
	shown in  Fig.\ref{DL1}(b). This  Fig.\ref{DL1}(b) confirms the order parameter by presenting 
	 a number of states equal to one in the case of synchronisation (dark blue). When the order parameter 
	is different from one, the synchronisation of all the systems in the network is not achieved. 
	At the moment, we can find partial synchronization called cluster which is characterized by a number of independent states of the network much less than the total number of elements. Based on
	Fig.\ref{DL1}(b), we see a thin region with cluster formation in the 
	area of transition from synchronization to desynchronization and vice versa.
	To better highlight the impact of the amplification ($C_2$) on the transition to
	synchronization, we have presented the order
	parameter of the multilayer network showing the transition to the synchronous
	state
	for three values of the amplification $C_2=0.5$, $C_2=1$ and
	$C_2=2$, showed in Fig.\ref{DL1} (c, d and e) respectively.
	Varying smoothly
	the intra and interlayer coupling for these three fixed values
	of $C_2$, we have obtained different
	areas of synchronization of the systems of the network. For these three figures the
	first
	domain (D1) represents the region where there is no
	synchronization in any layer and the second domain (D2)
	where the synchronization exists only in the first layer. The third domain (D3)
	represents the zone of synchronization of both layers and the last domain (D4)
	is showing the area where there is divergence between the states of the
	oscillators of the slave layer. Here, divergence means an infinite amplification of the amplitude 
	of the oscillations of the systems of  the slave layer leading to an explosion ($y^1, y^2, y^3$ tend 
	towards the inifinite). In summary, we notice that,
	the area of synchronization of both layers increases when the amplification
	coefficient decreases. So in Fig.\ref{DL1}(c) this zone
	of synchronization of both layers is the largest domain (D3) and the zone (D2) where
	only the
	master layer synchronizes is almost nonexistent. Therefore when $C_2$ increases,
	it takes
	stronger values of the coupling in the master layer for the slave layer to
	become synchronized as can be seen in Fig.\ref{MSF1}(b). Fig.\ref{DL} is used
	to confirm  and to present the transition to the synchronization using the 
	Pearson correlation \cite{pearson1920notes} (defined by Eq.\ref{Pearcor}) between the first variables 
	of each oscillator of both layers, 
	the time series of the first variables of the oscillators label 1 and 15 of both layers and the mean phase of the oscillators of the network.

	 	\begin{equation}
	 	\label{Pearcor}
	 	\rho(x^{1},y^{1}) = \frac{\sum\limits_{i = 1}^N (x_{i}^{1} - \bar{x}^{1})(y_{i}^{1} - \bar{y}^{1})}{\sqrt{\sum\limits_{i = 1}^N (x_{i}^{1} - \bar{x}^1)^{2}} \sqrt{\sum\limits_{i = 1}^N (y_{i}^{1} - \bar{y}^{1})^{2}}}\\
	 	 \end{equation}
	 	 where $\bar{x}^{1} = \sum\limits_{i = 1}^N x_{i}^{1}$ and $\bar{y}^{1} = \sum\limits_{i = 1}^N y_{i}^{1}$ are the mean of the states variables $x_{i}^1$ and $y_{i}^1$. 
	 	 In this manuscript, the colour yellow used in the correlation refers to the oscillators that synchronize and the colour blue to those that 
	 	 do not synchronize.

	Fig.\ref{DL}(a) shows the correlation between the first variables of 
	the oscillators
	of the master
	and slave layer for $C_0=10$ and $C_2=0.2$ (these values are taken almost at the
	border of
	the separation of the red and blue zone in Fig.3(a)). This figure presents
	yellow and blue colors to identify those oscillators of the multilayer network
	that synchronize and those who do not synchronize, respectively. In this
	figure, the x-axis corresponds to the index used to identify the oscillators of
	the master
	layer ($i$) and the y-axis corresponds to the index used to identify the
	different oscillators of the slave layer ($j$). According to this figure, the
	first 14  oscillators of both layers synchronize.  To present in detail the
	situation we plot in Fig.\ref{DL}(b) for the same fixed values of the parameters
	the mean phase for oscillators in both layers, where the labels 1 to 30 belong
	to the master layer and 30 to 60 to the slave layer. Here we notice two
	different groups in each layer: the synchronous group (first 14 oscillators of
	each layer) and the asynchronous group, so the transition to the synchronization
	is done by cluster formation. To show the effect of the amplification for these
	fixed values of parameters, we have in Fig.\ref{DL}(c) the time series of the
	variables for oscillators labels $1$,and $15$. This figure presents a chaotic
	evolution of these variables as a
	function of time as well as the amplification, these effects of 
	synchronization with
	amplification are well appreciated in Fig.\ref{DL}(d,f) for the oscillators
	indicated in the axes. According to Fig.\ref{DL}(a) the $15^{th}$ oscillator of
	the master layer and the $14^{th}$ oscillator of the slave layer can not
	synchronize (see Fig.\ref{DL}(e)).  At the end, to confirm the synchronization
	with amplification in the
	multilayer network, we present in Fig.\ref{DL}(g) the
	correlation between the elements of both layers for $C_0=10$ and $C_2=1.55$.
	This correlation shows the synchronization of the multilayer network which can
	be
	appreciated in Fig.\ref{DL}(h) where we plot the mean phase for oscillators in
	both layers. The time
	series
	of the oscillators $1$ and
	$15$ (Fig.\ref{DL}(i))
	show the amplification described in Eq.\ref{layer2}. We can appreciate the synchronization with 
	amplification of the
	multilayer network in the Fig.\ref{DL}(j,k,l) where
	complete synchronization of some chosen
	oscillators is shown.
\begin{figure}[htp!]
	\includegraphics[width=8.5cm, height=6cm]{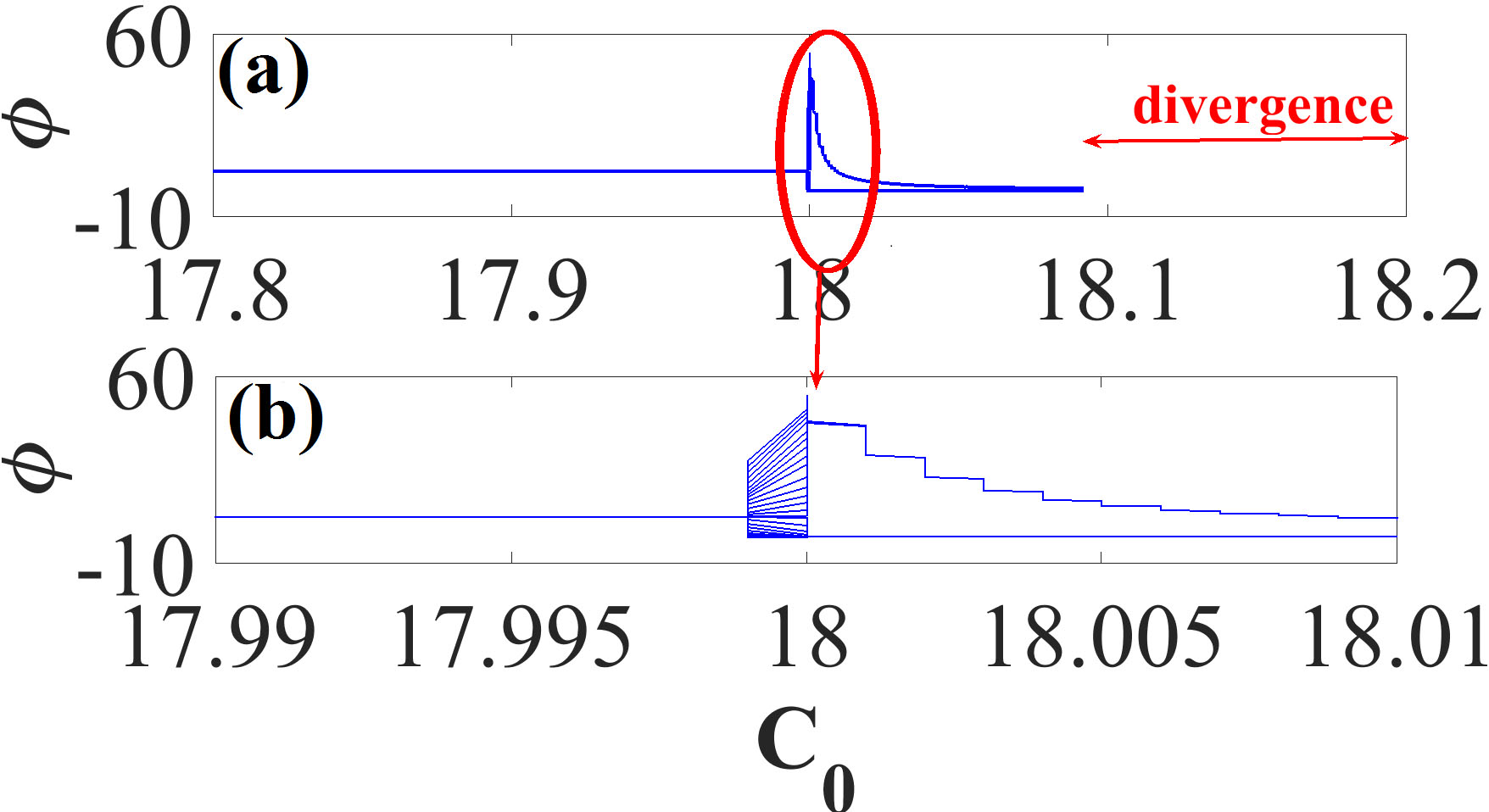}
	\includegraphics[width=8.5cm, height=4cm]{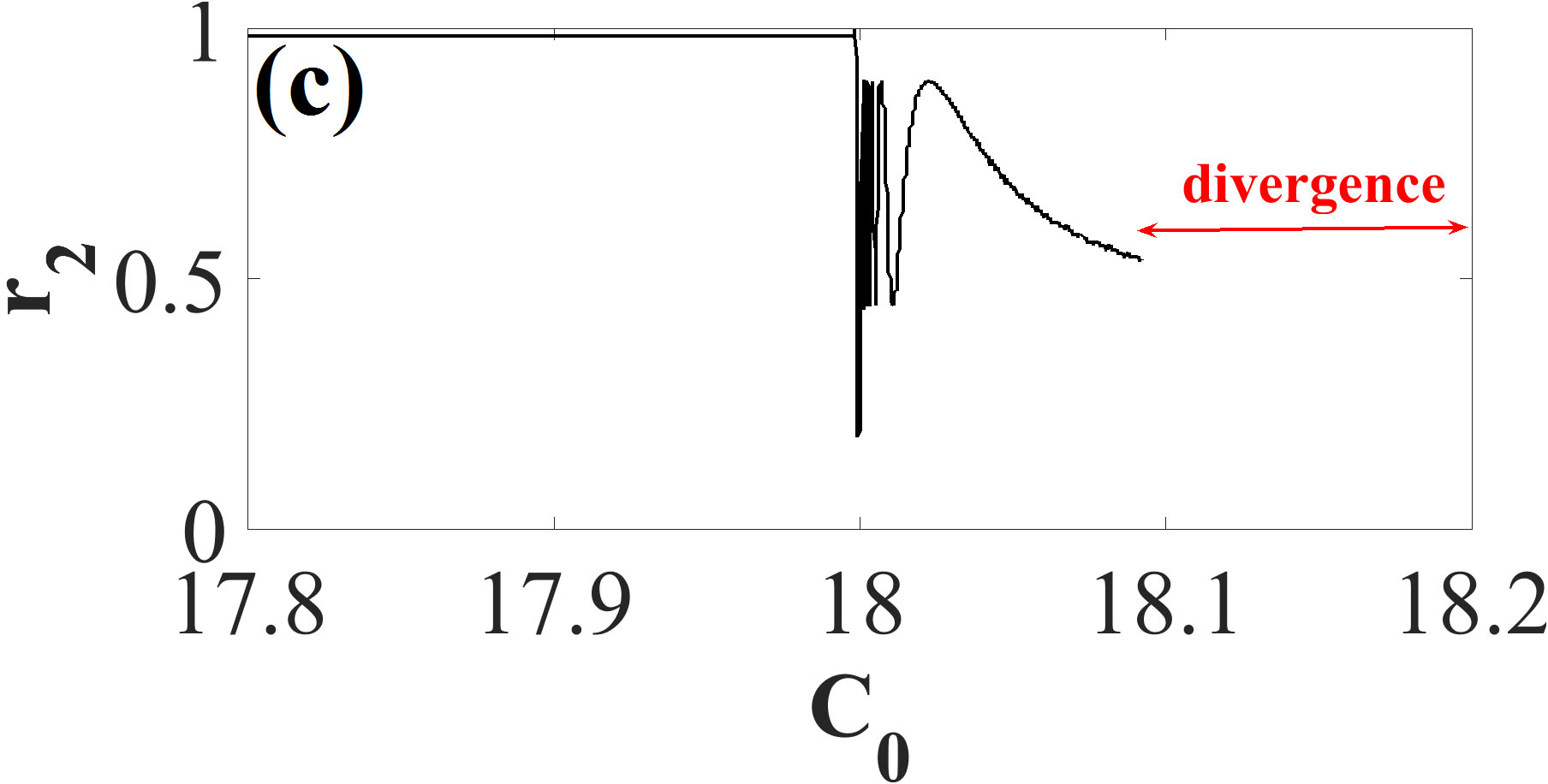}
	\includegraphics[width=8.5cm, height=4cm]{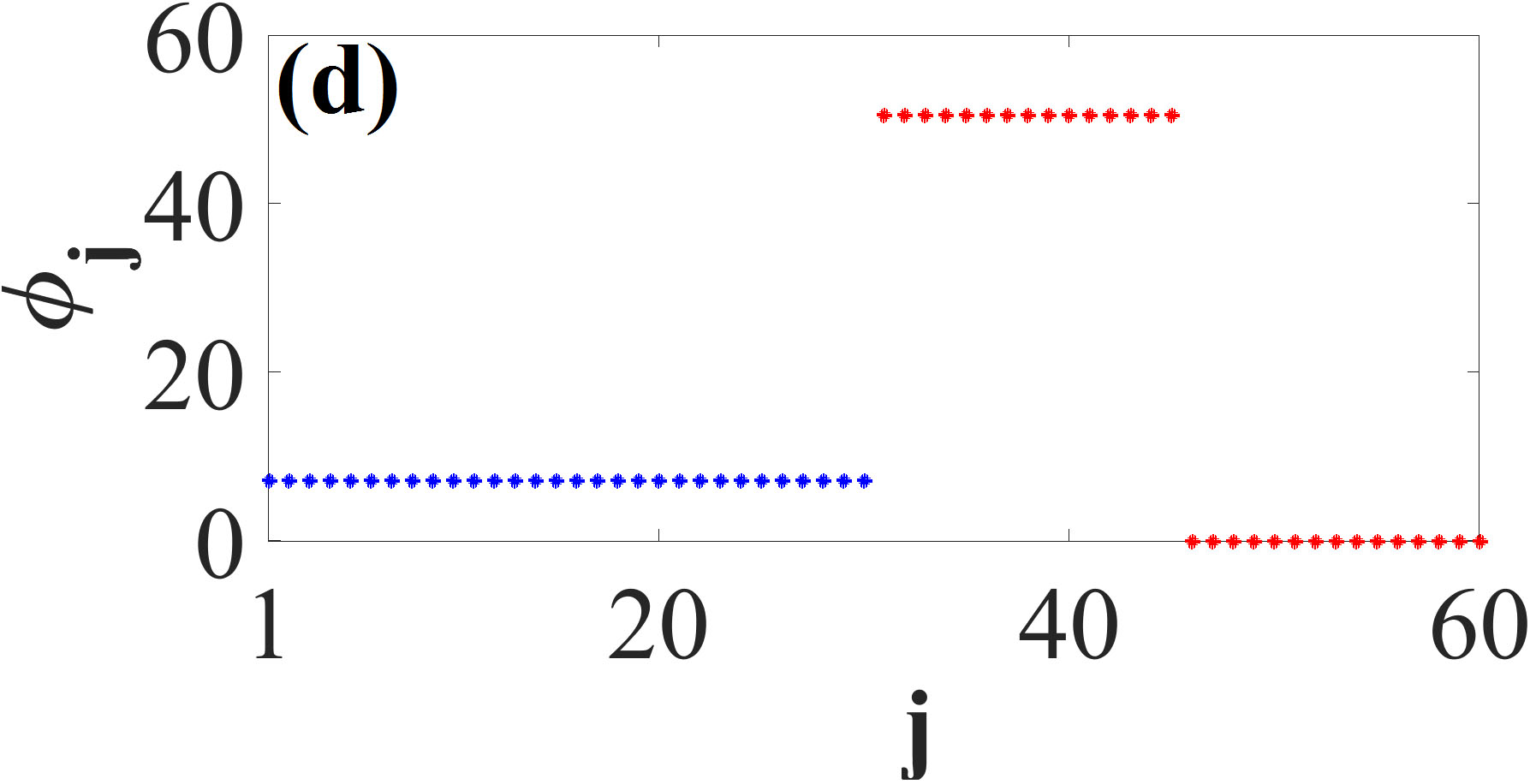}
	\includegraphics[width=9cm, height=4cm]{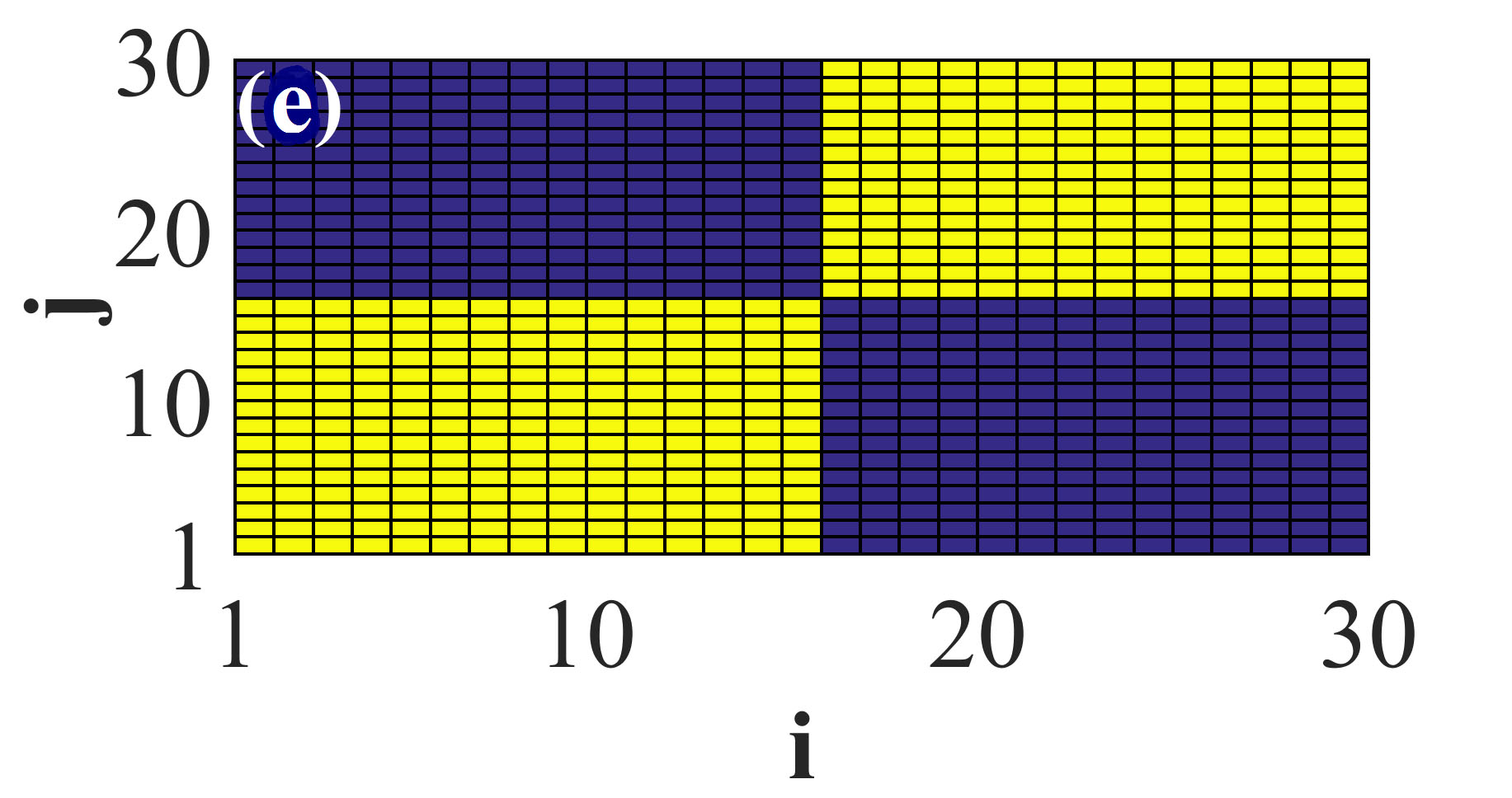}
	\caption{\footnotesize{Road to divergence: (a) Variation of the phases
			of the	slave layer for $C_2=2$, $\epsilon_1=0.065721$ and
			$\epsilon_2=10\epsilon_1$.	(b) Zoom of the variation of the phases of 
			the	slave layer for $C_2=2$, $\epsilon_1=0.065721$ and $\epsilon_2=10\epsilon_1$. 
			(c)) Order parameter of the slave layer for $C_2=2$, $\epsilon_1=0.065721$ 
			and	$\epsilon_2=10\epsilon_1$. (d) Mean phase of the oscillators of the 
			network and (e) correlation between the oscillators of the 
			slave layer for $\epsilon_1=0.065721$ and	
			$\epsilon_2=10\epsilon_1$, $C_2=2$ and $C_0=18$.  (The yellow colour indicates 
			the oscillators that synchronize and the blue colour those that do not synchronize). }}
	\label{DL11}
\end{figure}

	To appreciate the dynamics at the border of the separation of the domain D3 and
	D4 of the Fig.\ref{DL1}(e) we present in Fig.\ref{DL11}(a) the variation of the
	phase of all oscillators of the slave layer for $C_2=2$,
	$\epsilon_1=0.065721$, $\epsilon_2=10\epsilon_1$ and interlayer coupling ($C_0$)
	varying from 17.8 to 18.2. At this range of values of the interlayer coupling,
	the slave layer presents different dynamics. Before $C_0=18$
	this slave layer shows a synchronization of all oscillators of the slave layer. 
	After this synchronisation there follows a slight zone of 
	disturbance (the zoom is given at Fig.\ref{DL11}(b)) before the division into two groups which drives 
	the layer towards the divergence.  
	This abrupt
	change presented reminds of an explosive desynchronization \cite{ref18}. Given
	the form of the connection (unidirectionnal coupling) between the first and second layer, the oscillators of the slave layer can only remain synchronous and stable for a certain range of values of $C_0$.  This dynamic leads the slave layer to the
	divergence that we observe in
	Fig.\ref{DL11}(a) on the mean phase and in Fig.\ref{DL11}(c) with the order
	parameter. To understand more clearly the dynamics of the slave layer at this
	value of interlayer coupling where we
	have destruction of the synchronization, we show in
	Fig.\ref{DL11}(d,e)  the mean phase of the multilayer network and the
	correlation between the oscillators of the slave layer respectively. The mean
	phase presents a
	phase synchronization of all the oscillators of the first layer but in the slave
	layer we have two clusters formations. This cluster formation in the slave layer
	is confirmed using the correlation between the
	oscillators of the slave layer and then we
	can  appreciate the formation of the these two
	clusters by the yellow colour.

	\section{Applications to other systems}

	The behavior of a multilayer network shown in the previous Sections is not restricted to a system of R\"ossler oscillators, 
	it can also be obtained with other systems and topologies. In this Section we shall study a network of jerk oscillators and another of Liénard oscillators
	with different topologies and show that they reproduce the same behaviors.

	\subsection{Synchronization with amplification in a multilayer network of jerk oscillators}

	Here we investigate the dynamics of a multilayer network of jerk\footnote{Based on Ref.\cite{ref29}, a system is considered as jerk if the flow can be rewritten as a third order differential equation in a single scalar variable. For an isolated system, the jerk system can be defined as:
		$\dddot{x}_2 = \alpha \left(- \beta \left(- \dot {x}_1 + \dot {x}_2 -
		\gamma \ddot {x}_1 + \frac{\gamma}{\alpha}\ddot {x}_2\right) + \dddot {x}_1\right)$.
		where $\ddot{x}_1=I'(x_2))\dot{x}_2$ and $\dddot{x}_1=I''(x_2))\dot{x}_2^{2} + I'(x_2))\ddot{x}_2.$
		Then, $\dddot{x}_1$ is called jerk function.}  chaotic oscillators where the first layer is described by Eq.\ref{jerk1}.

	\begin{equation}
	\label{jerk1}
	\begin{array}{cc}
	\left\{ \begin{array}{l}
	\dot x_{i}^{1} =  I(x_{i}^{2}) +
	{\epsilon_{1}}(x_{i+1}^1 + x_{i-1}^1 - 2x_{i}^1),\\
	\dot x_{i}^3 = \alpha(-x_{i}^3 + I(x_{i}^2)),\\
	\dot x_{i}^3 = \beta(-x_{i}^2 + x_{i}^2-\gamma x_{i}^3).
	\end{array} \right.
	\end{array}
	\end{equation}\\
	The slave layer is described by Eq.\ref{jerk2}
	\begin{equation}
	\label{jerk2}
	\begin{array}{cc}
	\left\{ \begin{array}{l}
	\dot y_{j}^1 =  \frac{I(x_{j}^2)}{C_2} +	{\epsilon_{2}}(y_{j+1}^1 + y_{j-1}^1-2y_{j}^1) +
	C_0({x_{j}^1} - {C_2}{y_{j}^1}), \\
	\dot y_{j}^2 = \alpha(-y_{j}^3 + I(x_{j}^2)/C_2),\\
	\dot y_{j}^3 = \beta(-y_{j}^1 + y_{j}^2-\gamma y_{j}^3).
	\end{array} \right.
	\end{array}
	\end{equation}\\
	Where the piecewise linear function is
	\begin{equation}
	\label{Piecewise}
	I(x^2)=\left\{\begin{array}{ccc}
	-x^2 \quad {if} \quad  x^2\leq 1, \\
	-1 \quad{otherwise}.\quad
	\end{array}\right.
	\end{equation}
	$\alpha=0.025$, $\beta=0.765$, $\gamma=0.0938$ are the systems parameter.
	$\epsilon_1$ and $\epsilon_2$ are the intralayer coupling strength of the drive
	and slave layer of the network. The connection between the nodes of the same
	layer is bidirectional and they are arranged on a ring. Moreover, connection between the nodes of different 
	layers is unidirectional and it only concerns the oscillators with the same index in both layers.
	
	\begin{figure}[htp]	
		\includegraphics[width=9cm, height=6cm]{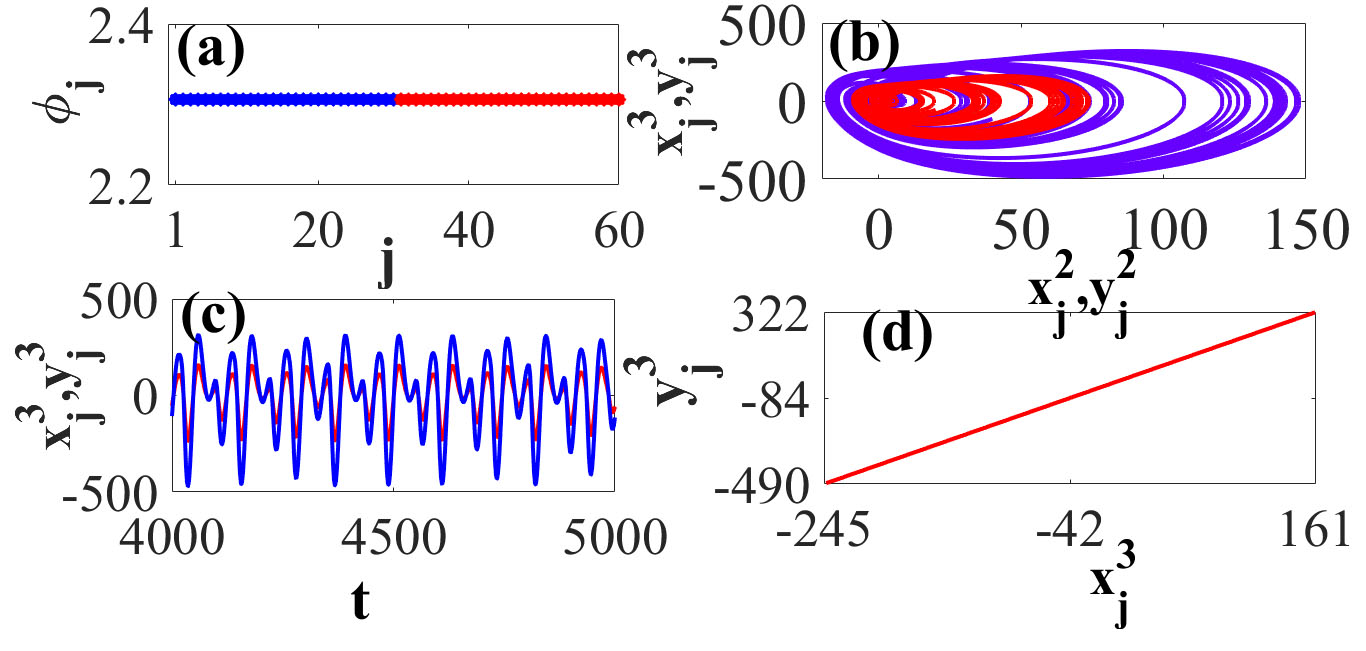}\\
		\includegraphics[width=9cm, height=6cm]{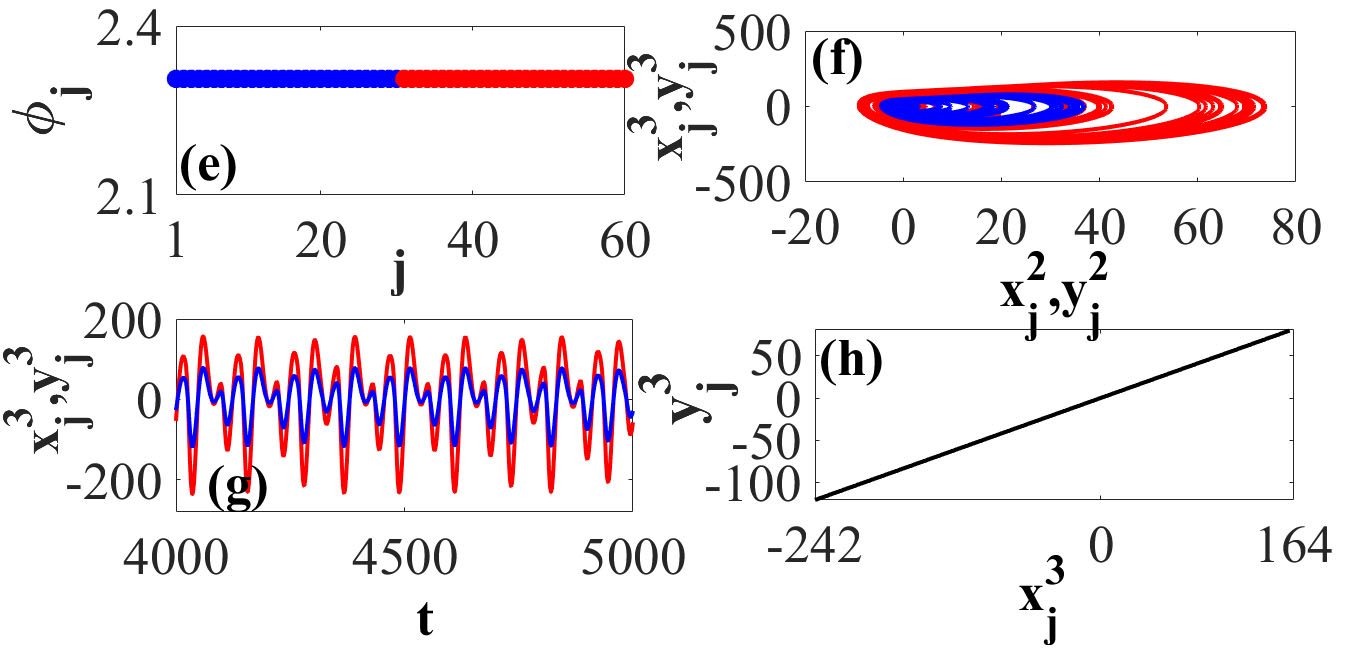}	
		\caption{\footnotesize{Dynamics of the two layers of the network of jerk oscillators,
				where the master layer is represented in red and the slave layer in blue.
				For $\epsilon_1=\epsilon_2=2,
				C_0=5$ and $C_2=0.5$: (a) mean phase of the first and second layer, (b) attractors of the
				oscillators in the master and slave layer, (c) time series of the oscillators of the
				master and slave layer and (d) synchronization between the oscillators of the first
				and second layer in the network. For
				$\epsilon_1=\epsilon_2=2, C_0=5$ and $C_2=2$: (e) mean phase of the first and second layer,
				(f) attractors of the oscillators in the master and slave layer, (g) time series of the
				oscillators of the first and second layer and (h) synchronization between the
				oscillators of the first and second layer in the network. }}
		\label{jerk}
	\end{figure}
	
	Eq.\ref{jerk1} and Eq.\ref{jerk2} were solved numerically considering $N=30$
	jerk oscillators per layer with the systems parameter defined above.
	Fig.\ref{jerk} is obtained for $\epsilon_1=2$, $\epsilon_2=2$ and $C_0=2$. Two
	values for the amplification coefficient are considered:
	$C_2=0.5$ and $C_2=2$. For $C_2=0.5$ we can clearly appreciate in Fig.\ref{jerk}(a) the complete phase synchronization of all
	oscillators in both layers. This phase synchronization is a
	major condition to obtain amplification in the systems of the network. As
	defined by our model, it emerges that for this values of $C_2$ the oscillators
	of the slave network are supposed to be amplified compared to oscillators of
	the master layer (see Fig.\ref{jerk}(b,c) where in red we show the oscillators of the master layer and in blue those of the slave layer). The
	amplification between master and slave layer is perfectly observed in
	Fig.\ref{jerk}(d). In the second case we consider $C_2=2$. As in the previous
	case we have phase synchronization between the oscillators of both layers (see
	Fig.\ref{jerk}(e)). For this value we have an amplification in the oscillators
	of the master layer (see Fig.\ref{jerk}(f,g)). In the same vein we present in
	Fig.\ref{jerk}(h) the synchronization between
	the oscillators in the first and second layer of the network.
	
	Therefore, interlayer synchronization can be obtained with amplification or
	reduction depending on the value of the coefficient $C_2$.

	\subsection{Chimera states with amplification in a multilayer network of
		Li\'enard system}

	Let us consider a network of Liénard systems expressed as in reference \cite{ref36}
	where the authors chose to investigate the dynamics of the oscillators basing
	themselves on an attractive and repulsive global coupling.
	In the same vein, in the Refs.\cite{ref30,ref31} the authors present some
	behaviour such as clusters, pattern formation, synchronization and so on
	according to the attractive and repulsive coupling.
	
	In this subsection, we consider the Li\'enard model with a intralayer topology
	defined as in \cite{ref36}.\\
	First or master layer

	\begin{equation}
	\label{lienard1}
	\begin{array}{cc}
	\left\{ \begin{array}{l}
	\dot{x}_{i}^1= x_{i}^2,\\
	\dot{x}_{i}^2 = -\alpha x_{i}^1 x_{i}^2 - \beta (x_{i}^1)^3 - \gamma x_{i}^1 +
	\displaystyle K\left[\left(\overline{x}^{2} - x_{i}^2\right) +
	\epsilon_1\left(\overline{x}^{1} - x_{i}^1\right)\right].
	\end{array} \right.
	\end{array}
	\end{equation}\\
	Where $ \overline{x}^1=\frac{1}{N}\sum\limits_{i = 1}^N x_{i}^1$ and  $ \overline{x}^2=\frac{1}{N}\sum\limits_{i = 1}^N x_{i}^2$.\\
	Second or slave layer
	
	\begin{equation}
	\label{lienard2}
	\begin{array}{cc}
	\left\{ \begin{array}{l}
	\dot{y}_{j}^1= y_{j}^2 + C_0({x_{i}^1} -	{C_2}{y_{j}^1}),\\
	\dot{y}_{j}^2= -\alpha \frac{x_{j}^1 x_{j}^2}{C_2} - \beta \frac{(x_{i}^1)^3}{C_2}  -
	\gamma y_{j}^1 + \displaystyle K\left[\left(\overline{y}^{2} - y_{j}^2\right) +
	\epsilon_2\left(\overline{y}^{1} - y_{j}^1\right)\right] \\
	+ C_0({x_{j}^2} - {C_2}{y_{j}^2}).
	\end{array} \right.
	\end{array}
	\end{equation}\\
	With $ \overline{y}^1=\frac{1}{N}\sum\limits_{i = 1}^N y_{i}^1$ and  $ \overline{y}^2=\frac{1}{N}\sum\limits_{i = 1}^N y_{i}^2$.

	The system's parameters $\alpha$, $\beta$ and $\gamma$ are  selected exactly as
	in \cite{ref36}, $K$ is the strength of coupling, $\epsilon_1$ and $\epsilon_2$
	are the intralayer global mean field coupling and $C_2$ is the amplification coefficient.
	The coupling between the nodes of different layers is unidirectional and 
	it only concerns the oscillators with the same index in both layers and $N=100$ Li\'enard systems.
	
	For the numerical simulation, we consider two cases, one where the amplifcation
	is less than one (the systems of the slave layer are amplified and the systems
	of master layer are reduced) and another when the amplification is  greather than one (the systems of the slave layer are reduced and the systems of master layer are
	amplified).

	\begin{figure}[htp!]
		\begin{center}
			\includegraphics[width=9cm, height=9cm]{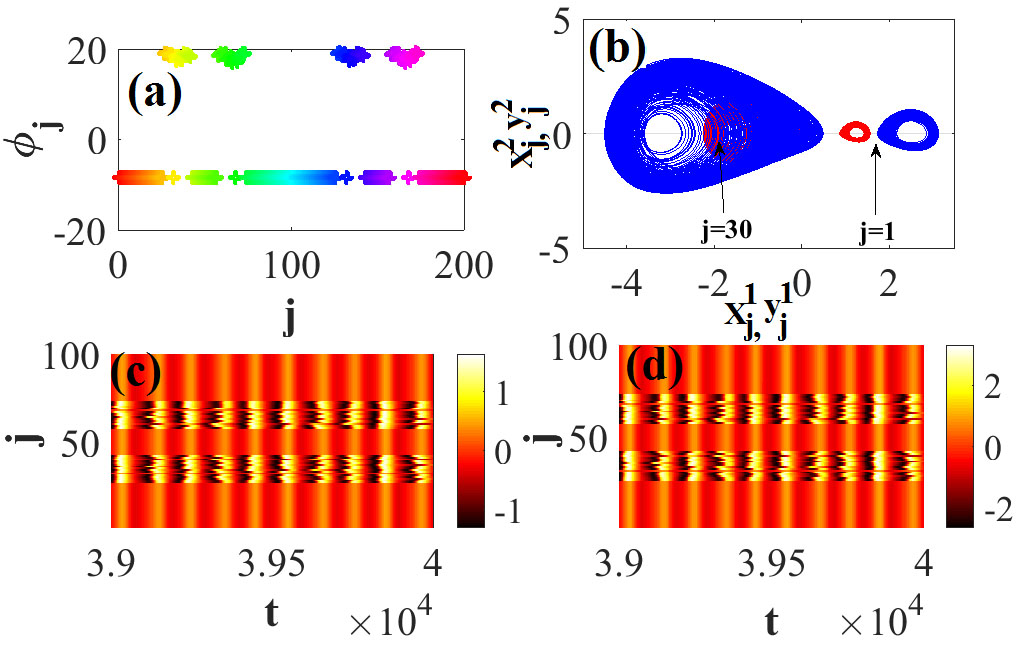}
			\includegraphics[width=9cm, height=9cm]{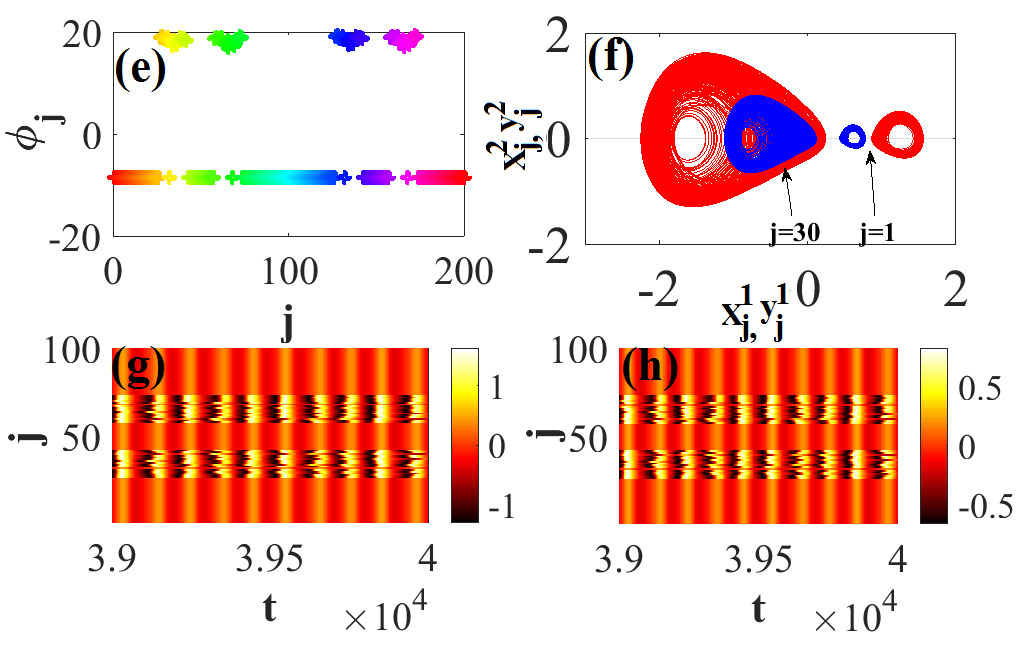}	
		\end{center}
		\caption{\footnotesize{Examples of behaviour
				of the multilayer network of
				Li\'enard systems. Amplification for $K=0.9$,
				$\epsilon_1=\epsilon_2=-0.57$, $C_2=0.5$: (a) Snapshot of all the oscillators of the multilayer network showing a multichimera state, (b) phase portrait of the oscillators $j=1$ and $j=30$ of both layers and (c,d) temporal dynamics of all the nodes in the first and second layer in the network.  Reduction for $K=0.9$,
				$\epsilon_1=\epsilon_2=-0.57$, $C_2=2$: (e) Snapshot of all the oscillators of the multilayer network showing a multichimera state, (f) phase portrait of the oscillators $j=1$ and $j=30$ of both layers and (g,h) temporal dynamics of all the nodes in the first and second layer in the network.}}
		\label{Lie1}
	\end{figure}

	Fig. \ref{Lie1} elaborates the Chimera-I states based on Refs.\cite{ref36,ref37}.
	Fig. \ref{Lie1}(a) presents a snapshot of all the oscillators of both layers of
	the network (the first 100 systems correspond to the master layer and the rest
	is for slave layer). This figure presents a multichimera state in both
	master and slave layer (see Fig. \ref{Lie1}(c,d)) and
	a phase synchronization of both layers for $K=0.9$,
	$\epsilon_1=\epsilon_2=-0.57$, and $C_2=0.5$. For the same parameters
	we show in
	Fig.\ref{Lie1}(b) the attractor of the system for oscillator $j=1$ in the synchronization state and
	the attractor for $j=30$ in the incoherent state for both layers of the network
	(attractor red corresponds to the master layer and attractor blue for the slave layer). At $C_2=0.5$ the systems of the slave layer are
	supposed to be amplified as in Fig.\ref{Lie1}(b). In the same vein we present in
	Fig.\ref{Lie1}(e,f,g,h) for $C_2=2$ the same behaviours as in
	Fig.\ref{Lie1}(a,b,c,d).  We observe the
	same behaviours except that {are the systems of the master layer which} are amplified as
	$X=2Y$.\\
	Based on these results, we conclude that this form of coupling can lead to a
	chimera or multichimera state with amplification or reduction depending on the value of the amplification coeficient.

	\begin{figure}[htp!]	
		\begin{center}
		\includegraphics[width=9cm, height=8cm]{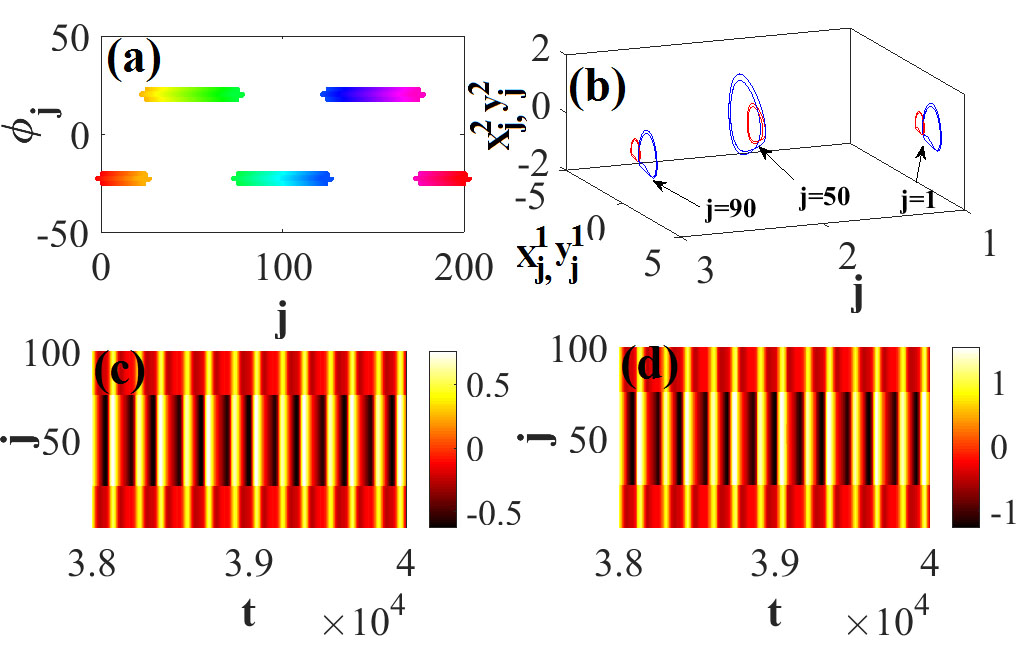}
		\includegraphics[width=9cm, height=8cm]{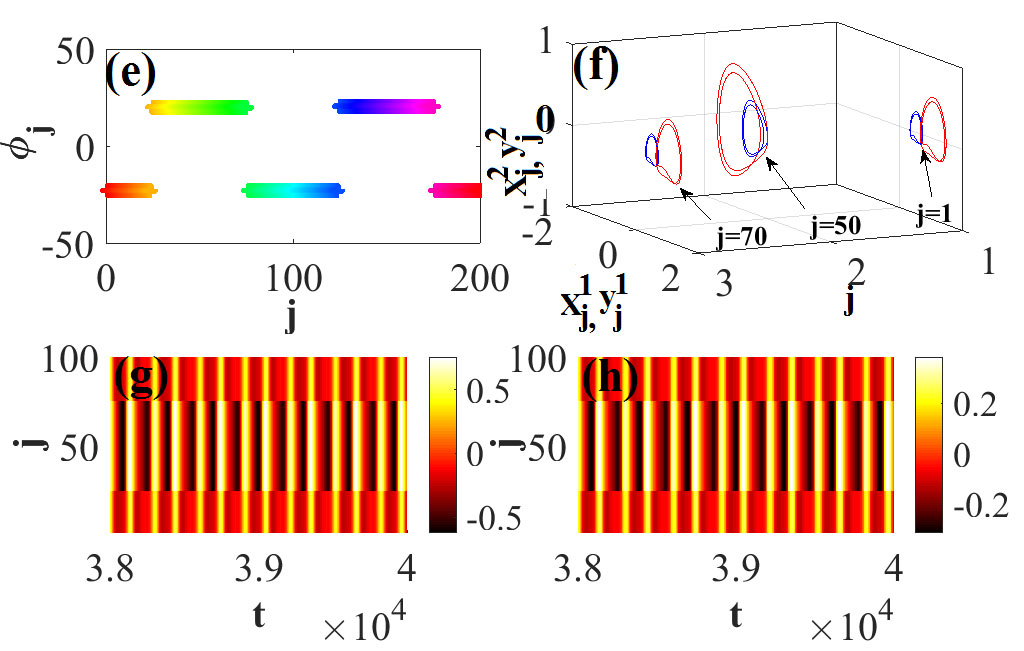}
		\end{center}	
		\caption{\footnotesize{Examples of behaviour
				of the multilayer network of
				Li\'enard systems. Amplification for $K=1.5$,
				$\epsilon_1=\epsilon_2=-0.1$, $C_2=0.5$: (a) Snapshot of all the oscillators of the multilayer network showing a cluster, (b) phase portrait of the oscillators $j=1$, $j=50$ and $j=90$ of both layers and (c,d) temporal dynamics of all the nodes in the first and second layer in the network.  Reduction for $K=1.5$,
				$\epsilon_1=\epsilon_2=-0.17$, $C_2=2$: (e) Snapshot of all the oscillators of
				the multilayer network showing a cluster, (f) phase portrait of the
				oscillators $j=1$, $j=50$ and $j=70$ of both layers and (g,h) temporal dynamics of all
				the nodes in the first and second layer in the network.}}
		\label{Lie2}
	\end{figure}

	Let us now consider the following parameters $K=1.5$ $\epsilon_1=\epsilon_2=-0.1$
	that lead to a cluster formation and we choose two values of amplification
	parameter: (a) $C_2=0.5$ (see Fig.\ref{Lie2}(a,b,c,d))
	and (b) $C_2=2$ (see Fig.\ref{Lie1}(e,f,g,h)). According to these
	two cases we notice that this cluster state can be maintained with an
	amplification or reduction based on the value of $C_2$.

	\section{Conclusion}
	\noindent
	Summarizing, we have studied and characterized numerically the synchronization
	and the amplification of signals in a multilayer network of R\"ossler, jerk or Liénard oscillators. Using tools for studying synchronization in the network such as
	master stability function, the order parameter, we have demonstrated that, the
	existence of synchronization in the second layer is conditioned by the first. To
	obtain synchronization of both layers at the same value of the intralayer
	coupling, the amplification coefficient must be sufficiently low (see
	Fig.\ref{MSF1}(b)). The key role of amplification is demonstrated by analyzing the order parameter of the first and second layer. This parameter leads the network to different dynamics such as cluster formation and synchronization.

	From a theoretical point of view, this work contributes to the
	advancement in the understanding of the phenomenon of synchronization and
	amplification of signals between two coupled networks. From a practical point of
	view, the results may be useful in many technological applications. For example, in several 
	mechanical systems, the transmission of movements or orders is done through a driving belt
	or gear \cite{denijs1997transmission}. Concerning this process, one of the most important parameters is the 
	transmission ratio, here represented by $C_2$, the amplification parameter. The amplification coefficient 
	could be the transmission ratio between gears or pulley-belt systems \cite{farshidianfar2014identification}.

 	\section{Data Availability Statement}
 	
 	The data that supports the findings of this study are available within this article.

	\section*{Acknowledgements}
	
	HAC thanks ICTP-SAIFR and FAPESP grant 2016/01343-7 for partial support. VC and
	FFF acknowledge that this study
	was financed in part by the Coordena\c{c}\~{a}o de Aperfei\c{c}oamento de
	Pessoal de N\'{i}vel Superior - Brasil (CAPES) - Finance Code 001.

	\appendix
	\section{}

	\textbf{ Synchronization conditions}\\

	We consider $X(x_{i}^{1}, x_{i}^{2}, x_{i}^{3})$ and $Y(y_{i}^{1}, y_{i}^{2}, y_{i}^{3})$ the state vector of the $i^{th}$ oscillator of the first and second layer respectively.\\
	 The dynamics of the first and second layers are given respectively by:
	 \begin{equation}
	 \label{Layer1}
	 \begin{array}{cc}
	 \left\{ \begin{array}{l}
	 \dot x_{i}^{1} =  - x_{i}^{2} - x_{i}^{3} +
	 {\epsilon_{1}}\sum\limits_{k = 1}^N A_{ik}^{1} (x_{k}^{1}
	 -x_{i}^{1}),\\
	 \dot x_{i}^{2} = x_{i}^{1} + ax_{i}^{2},\\
	 \dot x_{i}^{3} = bx_{i}^{1} + x_{i}^{3}(x_{i}^{1} - c).
	 \end{array} \right.
	 \end{array}
	 \end{equation}
	 and
	 \begin{equation}
	 \label{Layer2}
	 \begin{array}{cc}
	 \left\{ \begin{array}{l}
	 {{\dot y}_{j}^{1}} =  - {y_{j}^{2}} - {y_{j}^{3}} + {\epsilon_2}\sum\limits_{k = 1}^N A_{jk}^{2}
	 ({y_{k}^{1}} - {y_{j}^{1}}) + C_{0}(x_{j}^{1} - C_{2}y_{j}^{1}),  \\
	 {{\dot y}_{j}^{2}} = \frac{{x_{j}^{1}} + a{x_{j}^{2}}}{C_2},\\
	 {{\dot y}_{j}^{3}} = b{y_{j}^{1}} +
	 \frac{{{x_{j}^{3}}{x_{j}^{1}}}}{{{C_2}}} - c{y_{j}^{3}}.
	 \end{array} \right.
	 \end{array}
	 \end{equation}
	 Without any intralayer coupling $(\epsilon_1=\epsilon_2=0)$ both layers could synchronize with amplification 
	 depending on $C_2$.\\
	 We consider the synchronization error  $e=X - C_2Y$ with $\epsilon_1=\epsilon_2=0$.
	 So, the error dynamical system which is obtained only for the couple of systems with the same index in the first and second layers is given by Eq.\ref{Er1}:
	 \begin{equation}
	 \label{Er1}
	 \begin{array}{cc}
	 \left\{ \begin{array}{l}
	 \dot{e}_{i}^{1} = -e_{i} ^{2}- e_{i}^{3} - C_2C_0e_{i}^{1},\\
	 \dot{e}_{i}^{2}= 0,\\
	 \dot{e}_{i}^{3}= be_{i}^{1}   - ce_{i}^{3}.
	 \end{array} \right.
	 \end{array}
	 \end{equation}
	 To simplify the demonstration, we consider the Lyapunov functions $v_i (i=1,2,...,N)$ of system pairs $i$ and $j$ (with  $i=j$) of the first and second layer described as follow:
	 \begin{equation}
	 \label{Err2}
	 \begin{array}{cc}
	 v_i=\frac{1}{2} \left((e_{i}^{1})^{2} + (e_{i}^{2})^{2} + \frac{1}{b}(e_{i}^{3})^{2}\right)
	 \end{array}
	 \end{equation}
	 For any couple $(X,Y)$ we choose the following Lyapunov function candidate:
	 \begin{equation}
	 \label{Er2}
	 \begin{array}{cc}
	 v=\frac{1}{2} \left((e^{1})^{2} + (e^{2})^{2} + \frac{1}{b}(e^{3})^{2}\right).
	 \end{array}
	 \end{equation}
	 It is established (assuming $ a>0, b>0 $ and $c>0$) that the
	 system defined by Eq.\ref{Er1} is practically stable since
	 the time derivative of the Lyapunov function in Eq.\ref{Er2} is
	 bounded by a positive constant. This also means that the error
	 between the driver and response systems is sufficiently small,
	 but different from zero and could be considered as a tolerance
	 in the synchronization condition \cite{kakmeni2010practical}.
	 \begin{equation}
	 \label{Err3}
	 \begin{array}{cc}
	 \dot{v} \leq \frac{(e^{2})^{2}}{4c}.
	 \end{array}
	 \end{equation}
	 For the whole network, the Lyapunov function candidate $V$ can be defined as a sum of $v_i$:
	 \begin{equation}
	 \label{Er4}
	 \begin{array}{cc}
	 V = \sum\limits_{i = 1}^N v_i.
	 \end{array}
	 \end{equation}
	 And then for the whole network we can have:
	 \begin{equation}
	 \label{Err4}
	 \begin{array}{cc}
	 \dot{V} \leq \sum\limits_{i = 1}^N \frac{(e_{i}^{2})^{2}}{4c}.
	 \end{array}
	 \end{equation}
	 Based on Eq.\ref{Er1}, this boundedness is ensured by the fact that $e_{i}^{2}(t)$ is constant due to the fact that $\dot{e}_{i}^{2}(t)= 0$.
	 Thereby, from Eq.\ref{Err4} $e\rightarrow 0$, $X-C_2Y=0$ and induce $X=C_2Y$. we can obtain amplification or 
	 reduction depending on the value of the coefficient $C_2$.

	\section{}
	
	\textbf{Preliminaries to the Investigation of the Dynamics of the Network: the Master Stability Function}\\

	A regular problem that arises when analyzing the dynamics of a network is to find conditions that guarantee the synchronization of a 
	system of coupled identical nonlinear oscillators, so that all the oscillators converge asymptotically towards the same state.

	The Master Stability Function (MSF) developed by Pecora and Carroll \cite{ref9}, constitutes one of the most useful tools to analyze 
	the synchronization stability of a system of coupled identical nonlinear oscillators \cite{ref2,ref9,ref10}. We develop here only some 
	points of the principal idea of this method.
	
	Considering a network of $N$ identical coupled chaotic oscillators (or nodes), let $\mathbf{ x}_{i}$  a vector 
	with $m$ components necessary to describe the state of the $i^{th}$ node.  In general, in absence of any interaction 
	between the nodes 
	of the network, the evolution of a node is given by Eq.\ref{iso}:
		\begin{align}\label{iso}
			\mathbf{\dot x}_{i} = \mathbf{F(x}_{i}),
		\end{align}
		In this Eq.\ref{iso}, $\mathbf{F}$ is a function defined from $\mathbb{R}^{m}$ to $\mathbb{R}^{m}$ and is used to define 
		the local dynamics of the oscillators. To describe how the oscillators evolve when they are connected in a network, 
		we need to consider not only the local dynamics presented at Eq.\ref{iso}, but also how each node is affected by the ones 
		to which it is connected. So, the law governing the dynamical interaction of the $i^{th}$ node is defined as:
		\begin{align}\label{MS1}
			\mathbf{\dot x}_{i} = \mathbf{F(x}_{i}) + \sigma \sum\limits_{j = 1}^N \mathbf{G}_{ij} \mathbf{H(x}_{j}),
		\end{align}
		where $\sigma$ is a coupling strength, $H: \mathbb{R}^{m} \longrightarrow \mathbb{R}^{m}$ is an arbitrary output function of each node's variables using in the coupling. If we put the network in a synchronized state, we have $\mathbf{x}_{i}=\mathbf{s}$ for all nodes, where $\mathbf{s}$ is any m-dimensional vector. The only way all nodes have the same behavior is to have the sum, $\sum\limits_{j = 1}^N \mathbf{G}_{ij}$, be the same for all $i$. So, to obtain complete or identical synchronization the row sums of the coupling matrix must be the same for all rows.
		According to Pecora and Carrol \cite{ref9}, we can collect the node dynamical variables, functions and coupling in:
		\begin{subequations}\label{mSFFF}
			\begin{align}
				\mathbf{x} & = \left [\mathbf{x}_{1},\mathbf{x}_{2}, \ldots
				,\mathbf{x}_{N}
				\right], \\
				\mathbf{F(x)} & = \left[ \mathbf{F}(\mathbf{x}_{1}),
				\mathbf{F}(\mathbf{x}_{2}), \ldots, \mathbf{F}(\mathbf{x}_{N})
				\right], \\
				\mathbf{H(x)} & = \left[ \mathbf{H}(\mathbf{x}_{1}),
				\mathbf{H}(\mathbf{x}_{2}), \ldots, \mathbf{H}(\mathbf{x}_{N})
				\right],
			\end{align}
		\end{subequations}
		and $G$ be the matrix coupling coefficients $G_{ij}$, then based on Eq.\ref{mSFFF}(a,b and c), Eq.\ref{MS1} 
		can be written in a compact form as follow:

	\begin{align}\label{MSF}
		\mathbf{\dot x} = \mathbf{F(x)} + \sigma \mathbf{G} \otimes
		\mathbf{H(x)},
	\end{align}
	where $\otimes$ is the Kronecker product. So, according to the form of  Eq.\ref{MS1}, we can rewrite Eq.\ref{layer1} 
	in the same form with:

		\begin{equation}
		\label{Fx}
		\mathbf{F(x}_{i})\begin{array}{cc}
		\left\{ \begin{array}{l}
		\dot x_{i}^{1} =  - x_{i}^{2} - x_{i}^{3},\\
		\dot x_{i}^{2} = x_{i}^{1} + ax_{i}^{2},\\
		\dot x_{i}^{3} = bx_{i}^{1} + x_{i}^{3}(x_{i}^{1} - c).
		\end{array} \right.
		\end{array}
		\end{equation}\\
		According to \cite{ref9}, for an all-to-all coupling scheme, the connectivity matrix can be defined 
		as in matrix $G$. To couple the nodes of the layer we choose the $\mathbf{x}^{1}$ component 
		and then the matrix $H$ can be defined as in Eq.\ref{MG}:
	
	{\begin{align}\label{MG}
			\mathbf{H} =
			\begin{pmatrix}
				1 & 0 & 0 \\
				0 & 0 & 0 \\
				0 & 0 & 0
			\end{pmatrix},
			\;\;
			\mathbf{G} =
			\begin{pmatrix}
				1-N &  1 & \cdots & 1 & 1 &  1 \\
				1 & 1-N & \cdots &  1 & 1 &  1 \\
				\vdots & \vdots & \cdots & \vdots & \vdots & \vdots \\
				1 & 1 & \cdots & 1 & 1-N &  1 \\
				1 & 1 & \cdots & 1 & 1 & 1-N
			\end{pmatrix},
		\end{align}
		The Master Stability Function studies the stability of the global synchronization
		in the network. Therefore, the synchronous state is obtained when
		$\mathbf{x}_{1} = \mathbf{x}_{2}=\ldots = \mathbf{x}_{N} = \mathbf{s}$.

		Suppose our system is synchronized and we perturb it so that each node is, in general, 
		slightly “away” from the synchronized motion. 
			Let us consider $\xi_{i}$ a small perturbation of the $i^{th}$ node of the network. so that after 
			the perturbation $\mathbf{x}_{i} = \mathbf{s} + \xi_{i}$. For N oscillators of the first layer the
			collections of the variations can be expressed as $\xi = (\xi_{1}, \xi_{2},..., \xi_{N})$. Now we 
			can derive an equation of motion for the small perturbations that we will use to explore if 
			the synchronized state is unstable or stable.
			So, replacing the perturbation $\mathbf{x}_{i} = \mathbf{s} + \xi_{i}$ in Eq.\ref{MS1} and using 
			Taylor theorem expand of $\mathbf{F(s + \xi}_{i})$ and $\mathbf{H(s + \xi}_{i})$ to 
			first order (since $\xi_{i}$ is small) we have the following variational equation:
			\begin{align}\label{var1}
				\mathbf{\dot \xi}_{i} = \mathbf{DF(s) \xi}_{i} + \sigma \sum\limits_{j = 1}^N \mathbf{G}_{ij} \mathbf{DH(s) \xi}_{j},
			\end{align}
			Using tensor notation, we can write Eq.\ref{var1} in a more compact form:
		
		\begin{align}\label{VE}
			\mathbf{\dot \xi} = [\mathbf{1_{N}\otimes D\textbf{F(s)}} + \sigma
			\mathbf{G}
			\otimes \mathbf{DH(s)}]\xi,
		\end{align}
		where $\mathbf{1_{N}}$ is the identity matrix of order $N$, $\textbf{DF}$ and
		$\textbf{DH}$ are the $N\times N$ Jacobian matrices of the corresponding vector
		functions.
		
		The solution of Eq.\ref{VE} can be in the form $\xi_{i} \sim \exp \mu_{i}t$. 
		The exponents $\mu$ tell us if the perturbation grows ($\mu > 0$) or shrinks ($\mu < 0$), the 
		former indicating a direction that is unstable and the latter a stable direction. After 
		diagonalization of the second term of Eq.\ref{VE} we obtain the variational equations which are diagonal 
		in the node coordinates and are now uncoupled and individually given by:
		\begin{align}\label{VE1}
			\mathbf{\dot \xi_{k}} = [\textbf{DF(s)} + \sigma\alpha_{k}
			D\mathbf{H(s)}]\xi_{k}.
		\end{align}
		where $\alpha_{k}$ is an eigenvalue of $\textbf{G}$, $k=1,2,...N$. For each
		$k$,
		the form of each block of the Eq.\ref{VE} does not change, only the scalar
		multiplier $\sigma\alpha_{k}$ differs for each block.
		
		Therefore, these steps lead us to design the following master stability
		equation:
		\begin{align}\label{VE2}
			\mathbf{\dot \xi} = [\textbf{DF(s)} + \sigma\alpha \mathbf{DH(s)}]\xi.
		\end{align}

		Computing the Largest Lyapunov exponent of this Master
		Stability Equation Eq.\ref{VE2} we obtain what Pecora and Carroll called The
		Master Stability Function and therefore we achieve a stable synchronization
		state if the MSF turns negative \cite{ref9,reff9,refff9}.

	\section{}
	
	\textbf{ Calculation of the Order Parameter}\\

	Collective behavior of such an N-oscillator system is conveniently described by the order 
	parameter. The evaluation of this order parameter \cite{panaggio2015chimera} used the phase of each oscillator of the network. To define the phase let
	us consider an arbitrary signal $s(\tau)$ with time $\tau$ and 
	its Hilbert transformation to be $\tilde{s}(\tau)$, we have:
	\begin{equation}\label{AR1}
	\psi(\tau)=s(\tau)+i \tilde{s}(\tau)=R(\tau)\exp^{i\phi(\tau)},
	\end{equation}
	where $R(\tau)$ is the amplitude and $ \phi(\tau)$ the phase of the variable $s(\tau)$. If the instantaneous  phase is $ \phi_i(\tau)$, it can be determined through the following relation:
	\begin{equation}\label{AR2}
	\phi_i(\tau)=\tan^{-1}\left[\displaystyle\frac{\tilde{s}_i(\tau)}{s_i(\tau)}\right].
	\end{equation}
In this paper the calculation of the phase was carried out using in each case the variable that best describes the dynamics of the system.\\ Thus, from the expression of the phase $\phi_{i}$, the mean phase $\phi$  is an algebraic average calculated on the $N$ oscillators of the layer.
So, for a network of N oscillators the order parameter can be expressed as:
 \begin{equation}
 \label{ORDP}
 \begin{array}{cc}
  r = \frac{1}{N}\sum\limits_{i = 1}^N e^{j\phi_{i}}
 \end{array}
 \end{equation}
where $j^{2}=-1$, when $r \rightarrow 1$, phase synchronization is reached and when $r \approx 0$, the network is desynchronized.

\nocite{*}
\section*{References}
\bibliography{ref}

\end{document}